\documentstyle[11pt,newpasp,twoside,psfig]{article}
\def\edcomment#1{\iffalse\marginpar{\raggedright\sl#1\/}\else\relax\fi}
\reversemarginpar
\def    \beq       {\begin{equation}}
\def    \eeq       {\end{equation}}
\def    \abs       {{\rm abs}}
\def    \Angstrom  {\,{\rm\AA}}
\def    \micron    {\,\mu{\rm m}}
\def    \cm     {\,{\rm cm}}

\def	\g	{\,{\rm g}}

\def    \kms    {\,{\rm km\,s}^{-1}}
\def    \K      {\,{\rm K}}

\def    \ltsim  {\lesssim}      
\def    \gtsim  {\gtrsim}       

\def    \NH     {N_{\rm H}}
\def    \s      {\,{\rm s}}

\def	\mum	{\,\mu{\rm m}}
\def	\um	{\,\mu{\rm m}}
\def	\simali	{\sim\,}
\def	\lambdaeff {\lambda_{\rm eff}}
\def	\ppm    {\,{\rm ppm}}
\def	\nm     {\,{\rm nm}}
\def	\GHz    {\,{\rm GHz}}
\def	\eV     {\,{\rm eV}}
\def    \etapl  {\eta_{\rm PL}}
\def    \Nb	{M}
\def    \bT	{{\bf T}}
\def	\C	{{\rm C}}
\def	\rmH	{{\rm H}}

\def	\Eg	{E_{\rm g}}
\def	\bK	{{\bf \vec{k}}}
\def	\bKpht	{{\bf \vec{k}_{\rm phot}}}

\def	\nuphn	{\nu_{\rm phon}}
\def	\EQC	{E_{\rm QC}}
\def	\meff	{m^{\ast}}
\def	\meeff	{m_e^{\ast}}
\def	\mheff	{m_h^{\ast}}
\def    \Eth    {E_{\rm th}}
\def	\me	{m_e}
\def	\kB	{k_{\rm B}}

\def    \vdb    {{\rm vdB\,133}}
\def    \er     {\epsilon^{\prime}}
\def    \ei     {\epsilon^{\prime\prime}}
\def    \eb     {\epsilon_{b}}
\def    \ef     {\epsilon_{f}}

\def\lesssim{\mathrel{\hbox{\rlap{\hbox{\lower4pt\hbox{$\sim$}}}\hbox{$<$}}}}
\def\gtrsim{\mathrel{\hbox{\rlap{\hbox{\lower4pt\hbox{$\sim$}}}\hbox{$>$}}}}

\begin{document}
\title{Interaction of Nanoparticles with Radiation}
\vspace{-3mm}
\author{Aigen Li}
\affil{Theoretical Astrophysics Program, 
The University of Arizona,\\
Tucson, AZ 85721; {\sf agli@lpl.arizona.edu}}

\vspace{-2mm}
\begin{abstract}
Interstellar grains span a wide range of sizes
from a few angstroms to a few micrometers.
The presence of nanometer-sized or smaller particles
in the interstellar medium is indicated directly by
the interstellar far ultraviolet (UV) extinction,
the ubiquitous 3.3, 6.2, 7.7, 8.6, and 11.3$\mum$
polycyclic aromatic hydrocarbon (PAH) emission features,
the near and mid infrared broadband emission seen in
the IRAS 12 and 25$\mum$ bands and the COBE-DIRBE
3.5, 4.9, 12 and 25$\mum$ bands, the 10--100$\GHz$ 
Galactic foreground microwave emission,
and indirectly by the heating of interstellar gas.
For nanoparticles under interstellar conditions,
UV/visible photon absorption is the dominant
excitation process.
With a heat capacity smaller than or comparable to the energy
of an energetic stellar photon, nanoparticles are subject
to single-photon heating, followed by vibrational relaxation,
photoionization, and photodestruction.
With excited electrons spatially confined,
semiconductor nanoparticles are expected to 
luminesce efficiently.
This review focuses on the photophysics of nanoparticles 
with emphasis on the stochastic heating and the vibrational 
excitation of interstellar PAH molecules,
and the excitation of photoluminescence
with special attention given to silicon nanoparticles.
\end{abstract}

\vspace{-8mm}
\section{Introduction: Historical Perspectives of Studies of 
Interstellar Grain Compositions and Sizes}
When the existence of solid particles in interstellar space 
was first convincingly demonstrated by Trumpler (1930) 
through the discovery of color excess between the photographic
(with an effective wavelength $\lambdaeff$\,$\approx$\,$4300\Angstrom$) 
and visual ($\lambdaeff$\,$\approx$\,$5500\Angstrom$) magnitudes, 
the grains were estimated to be $\gtsim$$10^{-19}\g$, 
corresponding to radii of $\gtsim$$20\Angstrom$ (Trumpler 1930). 
By the end of the 1930s, a $\lambda^{-1}$ extinction law 
(i.e., the interstellar extinction varied approximately 
inversely with wavelength $\lambda$) in the wavelength range 
1--3$\mum^{-1}$ had been well established 
(Hall 1937; Stebbins, Huffer, \& Whitford 1939). 
At that time, metallic grains with a dominant size of 
$\simali$0.05$\mum$ (Schal\'{e}n 1936) or a power-law size 
distribution $dn(a)/da$\,$\simali$$a^{-3.6}$ in the size range
$80\Angstrom$\,$\ltsim$\,$a$\,$\ltsim$\,$1\cm$ (Greenstein 1938) 
were proposed to explain the $\lambda^{-1}$ extinction law,
partly because meteoritic particles (predominantly metallic) 
and interstellar grains were then thought to have the same origin.

In view of the spatial correlation between gas concentration 
and dust extinction, Lindblad (1935) suggested that interstellar
grains were formed by condensation from the interstellar gas
through random accretion of gas atoms. However, it was found 
later that in typical interstellar conditions, the Lindblad
condensation theory would result in a complete disappearance 
of all condensable gases and the grains would grow to sizes
($\simali$10$\mum$) well beyond those which would account for 
the interstellar extinction 
(see Oort \& van de Hulst 1946 and references therein).
By introducing a grain destruction process caused by grain-grain
collisions as a consequence of interstellar cloud encounters,
Oort \& van de Hulst (1946) further developed the interstellar
condensation idea and led to the ``dirty ice'' model
consisting of saturated molecules such as H$_2$O, CH$_4$,
and NH$_3$ with an equilibrium size distribution which could
be roughly approximated by a functional form    
$dn(a)/da\,$$\simali$$\exp\left[-5 \left(a/0.5\,\mu {\rm m}\right)^3\right]$
and an average size of $\simali$0.15$\mum$. 

The discovery of interstellar polarization 
(Hall 1949; Hiltner 1949) cast doubts on the ``dirty ice'' model 
since ice grains are very inefficient polarizers. 
This stimulated Cayrel \& Schatzman (1954) to 
consider graphite grains as an interstellar dust component 
since aligned graphite particles would be a very efficient 
interstellar polarizer because of their strong optical anisotropy.
The ``dirty ice'' model was also challenged by nondetection
of the 3.1$\mum$ feature of H$_2$O ice outside of molecular clouds
(Danielson, Woolf, \& Gaustad 1965; Knacke, Cudaback, \& Gaustad 1969).
This gave the late J. Mayo Greenberg (1922--2001) the incentive 
to perform the early experiments on the ultraviolet (UV) 
photoprocessing of low temperature mixtures of volatile 
molecules simulating the ``original'' dirty ice grains 
(Greenberg et al.\ 1972) to understand how and why 
the predicted H$_2$O was not clearly present. 
From such experiments was predicted a new component 
of interstellar dust in the form of complex organics
known as ``organic refractories'', containing 
a mixture of aliphatic and aromatic 
carbonaceous molecules (Greenberg et al.\ 2000). 

As an alternative to the interstellar condensation process,
Hoyle \& Wickramasinghe (1962) proposed that graphite 
particles with radii a few times 0.01$\mum$ could form in 
the atmospheres of cool N-type carbon stars, 
and subsequently be ejected into interstellar space 
by the stellar radiation pressure, 
and be responsible for (part of) the interstellar extinction 
and polarization. The graphite proposal further gained its strength 
from the prominent 2175$\Angstrom$ interstellar extinction hump 
detected by the {\it Aerobee} rocket observation (Stecher 1965),
which was generally attributed to small graphitic grains 
(Stecher \& Donn 1965), although its exact nature still 
remains unidentified (Draine 1989, 2003a).

Similarly, Kamijo (1963) first proposed that SiO$_2$ whose size
is of the order of $\simali$20$\Angstrom$, condensed in the atmospheres 
of cool M-type stars and expelled into interstellar space, 
could serve as condensation nuclei for the formation of ``dirty ices''.
It was later shown by Gilman (1969) that grains around oxygen-rich 
cool giants are mainly silicates such as Al$_2$SiO$_3$ and 
Mg$_2$SiO$_4$. Interstellar silicates were first detected 
in emission 
in the Trapezium region of the Orion Nebula (Stein \& Gillett 1969); 
in absorption toward the Galactic Center 
(Hackwell, Gehrz, \& Woolf 1970), 
and toward the Becklin-Neugebauer object 
and the Kleinmann-Low Nebula (Gillett \& Forrest 1973). 
Silicates are now known to be an ubiquitous
interstellar dust component (see \S1 in Li \& Draine 2001a for a review).

In the 1960s and early 1970s,
the extension of the extinction curve toward the middle 
and far UV ($\lambda$$\gtsim$$3\mum^{-1}$) was made possible 
by rocket and satellite observations,
including the rocket-based photoelectric 
photometry at $\lambda=2600\Angstrom$ and 2200$\Angstrom$
(Boggess \& Borgman 1964); the {\it Aerobee} Rocket spectrophotometry
at 1200$\Angstrom$\,$\ltsim$\,$\lambda$\,$\ltsim$3000$\Angstrom$ 
(Stecher 1965);
the {\it Orbiting Astronomical Satellite} 
(OAO-2) spectrophotometry at 
1100$\Angstrom$\,$\ltsim$\,$\lambda$\,$\ltsim$\,$3600\Angstrom$ 
(Bless \& Savage 1972);
and the {\it Copernicus} Satellite spectrophotometry at
1000$\Angstrom$\,$\ltsim$\,$\lambda$\,$\ltsim$\,$1200\Angstrom$ 
(York et al.\ 1973).
By 1973, the interstellar extinction curve had been 
determined over the whole wavelength range from 
0.2$\mum^{-1}$ to 10$\mum^{-1}$.
The fact that the extinction continues to increase 
in the far UV (e.g., see York et al.\ 1973)
implies that no single grain type with either a single size or 
a continuous size distribution could account for the observed 
optical to far-UV interstellar extinction (Greenberg 1973). 
This led to the abandonment of any one-component grain models 
and stimulated the emergence of various kinds of models 
consisting of multiple dust constituents,
including silicate, silicon carbide, 
iron, iron oxide, graphite, dirty ice, solid H$_2$, etc.\\[1mm]

In 1956 Platt first suggested that very small grains or 
large molecules of less than 10$\Angstrom$ in radius grown by 
random accretion from the interstellar gas could be responsible 
for the observed interstellar extinction and polarization. 
Platt (1956) postulated these ``Platt'' particles as 
quantum-mechanical particles containing many ions and 
free radicals with unfilled electronic energy bands.
Donn (1968) further proposed that polycyclic aromatic 
hydrocarbon-like ``Platt particles'' may be responsible 
for the UV interstellar extinction.

Greenberg (1968) first pointed out that very small grains with 
a heat content smaller than or comparable to the energy of a single 
stellar photon, cannot be characterized by a steady-state temperature
but rather are subject to substantial temporal fluctuations 
in temperature. Under interstellar conditions, grains larger than 
a few hundred angstroms in radius would attain an equilibrium 
temperature in the range of 15--25$\K$ 
(Greenberg 1968, 1971; Mathis, Mezger,
\& Panagia 1983; Draine \& Lee 1984; Li \& Draine 2001b). 
This temperature was long thought to be $\simali$3$\K$ 
before van de Hulst (1949) made the first realistic estimation 
of 10--20$\K$ for both metallic and dielectric grains.

In a study of the scattering properties of interstellar dust 
(albedo and phase function) determined from the 
OAO-2 observations at 
1500$\Angstrom$\,$\ltsim$\,$\lambda$\,$\ltsim$\,$4250\Angstrom$ 
of the diffuse Galactic light (Witt \& Lillie 1973), 
Witt (1973) first explicitly suggested 
a bi-modal size distribution for interstellar grains: 
large grains with radii $\gtsim$$2500\Angstrom$ would provide 
extinction in the visible region including scattering 
which is strongly forward directed, 
and small particles with radii $\ltsim$$250\Angstrom$ would 
dominate the UV region and contribute nearly isotopic scattering.

The presence of ``Platt'' particles 
in a dust cloud near M17 was first suggested 
by Andriesse (1978) based on an analysis of 
its IR spectral energy distribution 
which was shown to be characterized by 
a combination of widely different color temperatures 
(e.g., the far-IR and near-IR color temperatures are 
$\simali$38$\K$ and $\simali$150$\K$, respectively)
and that the 8--20$\mum$ emission spectrum is similar 
over a distance of $\simali$2$^{\prime}$ through the source.
Andriesse (1978) argued that the color temperature differences 
cannot be easily ascribed to a spatial variation of the dust 
temperature in this cloud, because the cloud profile appears 
to be independent of the wavelength; 
instead, ``Platt'' particles with sizes of $\simali$10$\Angstrom$ 
exhibiting temperature fluctuations up to $\simali$150$\K$ 
plus large grains with an equilibrium temperature of $\simali$36$\K$ 
can explain  the observed IR spectrum of M17.

In a near-IR (1.25--4.8$\mum$) photometric and spectrophotometric 
study of three visual reflection nebulae NGC 7023, 2023, and 2068, 
Sellgren, Werner, \& Dinerstein (1983) discovered that each nebula
has extended near-IR emission consisting of emission features 
at 3.3 and 3.4$\mum$ and a smooth continuum which can be described 
by a greybody with a color temperature $\simali$1000$\K$. They found
that the emission spectrum 
(i.e., the color temperature for the continuum 
and the 3.3$\mum$ feature) 
shows very little variation from source to source 
and within a given source with distance 
from the central star. Sellgren (1984) argued that 
this emission could not be explained by thermal emission 
from dust in radiative equilibrium with the central star 
since otherwise the color temperature of this emission 
should fall off rapidly with distance from the illuminating 
star; instead, she proposed that this emission is emitted by 
ultrasmall grains, $\simali$10$\Angstrom$ in radius, 
which undergo large excursions in temperature due to 
stochastic heating by single stellar photons. 

The presence of a population of ultrasmall grains 
in the diffuse ISM was explicitly indicated by 
the 12$\mum$ and 25$\mum$ ``cirrus'' emission 
detected by the {\it Infrared Astronomical Satellite} (IRAS) 
(Boulanger \& P\'{e}rault 1988), which is far in excess 
(by several orders of magnitude) of what would be expected 
from large grains of 15--25$\K$ in thermal equilibrium with 
the general interstellar radiation field. Subsequent measurements 
by the {\it Diffuse Infrared Background Experiment} (DIRBE) 
instrument on the {\it Cosmic Background Explorer} (COBE) satellite 
confirmed this and detected additional broadband emission at 
3.5$\mum$ and 4.9$\mum$ (Arendt et al.\ 1998). 

More recently, spectrometers aboard 
the {\it Infrared Telescope in Space} (IRTS) 
(Onaka et al.\ 1996; Tanaka et al.\ 1996) 
and the {\it Infrared Space Observatory} (ISO) 
(Mattila et al.\ 1996) have shown that the diffuse ISM 
radiates strongly in emission features at 3.3, 6.2, 7.7, 8.6, 
and 11.3$\mum$. These emission features, 
first seen in the spectra of the planetary nebulae NGC 7027 
and BD+30$^{\rm o}$3639 (Gillett, Forrest, \& Merrill 1973), 
have been observed in a wide range of astronomical environments 
including planetary nebulae, protoplanetary nebulae, 
reflection nebulae, HII regions, circumstellar envelopes, and
external galaxies (see Tielens et al.\ 2000 for a review). 
Often referred to as ``unidentified infrared'' (UIR) bands, 
these emission features are now usually attributed to free-flying
PAH molecules which are vibrationally excited upon absorption of 
a single UV/visible photon 
(L\'{e}ger \& Puget 1984; Allamandola, Tielens, \& Barker 1985;
Allamandola, Hudgins, \& Sandford 1999; 
Draine \& Li 2001; Li \& Draine 2001b) 
although other carriers have also been proposed such as 
carbonaceous grains with a partly aromatic character
(e.g., HAC [Duley \& Williams 1981, Duley, Jones, \& Williams 1990],
quenched carbonaceous composite [QCC; Sakata et al.\ 1992], 
coal [Papoular et al.\ 1996], fullerenes [C$_{60}$; Webster 1993]),
and surface-graphitized nanodiamonds 
(Jones \& d'Hendecourt 2000).

\vspace{3mm} 
Since the late 1970s, various modern interstellar dust models
have been developed (see Li \& Greenberg 2003, Draine 2004, 
Dwek et al.\ 2004 for recent reviews).
In general, these models fall into three broad categories:
\vspace{-1.0mm}
\begin{itemize}
\item \underline{The {\bf silicate-graphite} model} 
      -- this model consists of two (physically) separate 
         dust components each with a power-law size distribution 
         $dn(a)/da \sim a^{-3.5}$ in the size range 
         $50\Angstrom$\,$\ltsim$\,$a$\,$\ltsim$\,$0.25\mum$
         (Mathis, Rumpl, \& Nordsieck 1977 [hereafter MRN]; 
          Draine \& Lee 1984). 
         Modifications to this model were later made by 
         Draine \& Anderson (1985), Weiland et al.\ (1986), 
         Sorrell (1990), Siebenmorgen \& Kr\"{u}gel (1992), 
         Rowan-Robinson (1992), Kim, Martin, \& Hendry (1994),
         Dwek et al.\ (1997), Clayton et al.\ (2003),
         and Zubko, Dwek, \& Arendt (2003) 
         by including new dust components (e.g., amorphous carbon, 
         carbonaceous organic refractory, and PAHs) 
         and adjusting dust sizes (e.g., deriving dust size
         distributions using the ``Maximum Entropy Method''
         or the ``Method of Regularization'' rather than
         presuming a certain functional form).
         
         Recent developments were made by Draine and his coworkers 
         (Li \& Draine 2001b, 2002a, 2002b; Weingartner \& Draine 2001a) 
         who have extended the silicate-graphite grain model 
         to explicitly include a PAH component as 
         the small-size end of the carbonaceous grain population.
         The PAH component, containing $\approx 45\times 10^{-6}$
         of C relative to H, is represented by 
         a log-normal size distribution $dn(a)/d\ln a \sim 
         \exp\left\{-\left[\ln\left(a/a_0\right)\right]^2/
         \left(2\sigma^2\right)\right\}$ 
         with $a_0\approx 3.5\Angstrom$ and
         $\sigma\approx 0.4$ for $a$$\gtsim$$3.5\Angstrom$ 
         (see Li \& Draine 2001b). Note that with the PAH component
         added, the size distribution of the larger grains is changed
         from the simple MRN power-law (see Weingartner \& Draine 2001a;
         Zubko et al.\ 2003).
\vspace{-1.0mm}
\item \underline{The {\bf silicate core-carbonaceous mantle} model} 
      -- originally proposed by Greenberg (1978)  
      in the context of a complex cyclic evolutionary scenario 
      (see Greenberg \& Li 1999), 
      this model consists of larger silicate grains
      coated by a layer of carbonaceous organic refractory material,
      produced by UV photolysis of ice mixtures which attempts to
      simulate the physical and chemical processes occurring 
      in interstellar space. The most recent development
      of this model was that of Li \& Greenberg (1997), who modeled
      the core-mantle grains as 2:1 (the ratio of the length to the
      diameter) finite cylinders 
      (to account for the interstellar polarization)
      with a Gaussian size distribution for the mantle
      $dn(a)/da \sim
      \exp\left[-5\left(a-a_{\rm c}\right)^2/a_{\rm i}^2\right]$
      and a single silicate core radius $a_{\rm c}$$\approx$$0.07\mum$   
      and a ``cut-off'' size $a_{\rm i}$$\approx$$0.066\mum$.
      In addition, a PAH component and a population of 
      small graphitic grains are added respectively to 
      account for the far-UV extinction rise 
      plus the ``UIR'' emission bands  
      and the 2175$\Angstrom$ extinction hump.

      Again, modifications to this model were also made by considering 
      different coating materials (e.g., amorphous carbon, HAC), 
      including new dust type (e.g., iron, small bare silicates),
      and varying dust size distributions 
      (Chlewicki \& Laureijs 1988; Duley, Jones, \& Williams 1989;
      D\'{e}sert, Boulanger, \& Puget 1990; Li \& Greenberg 1998;
      Zubko 1999a). 
      The Duley et al.\ (1990) model exclusively 
      consists of silicate core-HAC mantle grains with a bimodal
      size distributions: (1) $\simali$43\% of the total dust mass in
      grains $\ltsim$$100\Angstrom$ which produce the far-UV extinction
      rise and the 2175$\Angstrom$ hump through an electronic transition
      of the OH$^{-}$ ions in low-coordination sites on or within 
      silicate grains, and (2) the remaining dust mass 
      in grains 0.05$\ltsim$$a$$\ltsim$$0.25\mum$ 
      with a power-law $dn(a)/da \sim a^{-3.5}$ size distribution.
      The ``UIR'' bands were attributed to the thermally isolated 
      aromatic ``islands'' of HAC material -- Duley et al.\ (1989)
      postulated that an absorbed stellar photon might remain
      localized in a single aromatic unit of the grain long enough
      for the unit to cool by IR vibrational relaxation rather than
      by transferring the photon energy 
      to the phonon spectrum of the grain. 
\item \underline{The {\bf composite grain} model} 
      -- realizing that grain shattering due to grain-grain 
      collisions and subsequent reassembly through agglomeration 
      of grain fragments may be important in the ISM, 
      Mathis \& Whiffen (1989) modeled the interstellar grains 
      as composite collections of small silicates, vacuum 
      ($\approx 80\%$ in volume), and carbon of various kinds 
      (amorphous carbon, HAC, organic refractories) with
      a power-law size distribution $dn(a)/da$$\simali$$a^{-3.7}$
      in the size range $0.03\mum$\,$\ltsim$\,$a$\,$\ltsim$\,$0.9\mum$.
      In addition, a separate small graphite component 
      containing $\approx 59\times 10^{-6}$ of C is needed
      to account for the 2175$\Angstrom$ extinction hump.
      Wright (1987) argued that a fractal structure would 
      be expected for interstellar grains formed through 
      the coagulation of small grain fragments created 
      from grain disruption caused by supernova shock waves.
      In view that the relative abundances of refractory elements
      in the ISM may be as low as 65\% of solar (Snow \& Witt 1996),
      Mathis (1996) revised the composite model in order to 
      satisfy the tighter abundance constraints. To optimize the use
      of the heavy elements, Mathis (1996) derived a vacuum fraction 
      of $\simali$45\%. The ``UIR'' emission, which remained unaccounted
      for in the models of Mathis \& Whiffen (1989) and Mathis (1996),
      was taken into account in Zubko et al.\ (2003) by including
      a population of PAH molecules.      
\end{itemize}
 
It is fair to say that, as can be seen from the above summarizing 
description of the key contemporary dust models, a consensus 
on the interstellar dust compositions and sizes is now 
approaching among various grain models, although the debate 
in details is still going on and is not expected to disappear 
in the near future (except that the IR spectropolarimetric 
observation at 3.4$\mum$ and 9.7$\mum$ may allow a direct test
of the core-mantle model for interstellar dust; 
see Li \& Greenberg 2002) ---
\begin{itemize}
\item Regarding the grain chemical composition, 
      the most generally accepted view is that interstellar grains 
      consist of amorphous silicates and some form of 
      carbonaceous materials; the former is inferred from 
      the 9.7$\mum$ Si--O stretching mode and 18$\mum$ O-Si-O bending 
      mode absorption features in interstellar regions as well as 
      the fact that the cosmically abundant heavy elements such as 
      Si, Fe, Mg are highly depleted; the latter is mainly inferred 
      from the 2175$\Angstrom$ extinction hump 
      (and the ubiquitous 3.4$\mum$ C--H stretching vibrational band) 
      and the fact that silicates alone are not able to provide 
      enough extinction. For the carbonaceous component, a wide range
      of dust materials have been suggested including amorphous carbon, 
      coal, C$_{60}$, diamond, graphite, HAC, PAHs, organic refractory, 
      and QCC (see Pendleton 2004 for a review of the carbonaceous
      component).  
\item It is also generally accepted that, through the analysis
      of the wavelength-dependent interstellar extinction and 
      polarization curves as well as the near and mid IR emission, 
      interstellar grain sizes may be separated into 
      two domains -- (1) the ``large'' grain component 
      (with radii $a$$>$$0.025\mum$; including the ``classical'' 
      grains [with $a$$\gtsim$$0.1\mum$]) 
      which is primarily responsible for the extinction, 
      polarization and scattering at visible wavelengths and 
      the IR emission at $\lambda$$\gtsim$$60\mum$;
      and (2) the ``very small grain'' component 
      (with $a$$<$$0.025\mum$) which contributes importantly 
      to the extinction in the vacuum-UV
      and emit strongly in the near and mid IR 
      at $\lambda$$\ltsim$$60\mum$ 
      (see Figs.\,8,16 of Li \& Draine 2001b)
      when transiently heated to high temperatures
      during quantized absorption events.

      While the size distribution for the ``classical grain'' 
      component is relatively well constrained by fitting 
      the observed interstellar extinction curve for an 
      assumed dust composition, our knowledge of 
      the size distribution $dn/da$ for the ``very small grain'' 
      component is better constrained
      by the interstellar near and mid IR emission
      due to the fact that, for $\lambda$$\gtsim$$0.1\mum$, 
      these very small grains are in the Rayleigh limit
      ($x \equiv 2\pi a/\lambda$$\ll$1) 
      and their extinction cross sections $C_{\rm ext}(a,\lambda)$
      per unit volume $V$ are independent of size, 
      so that the observed UV/far-UV 
      extinction curve $A(\lambda)/\NH$
      (${\rm mag\,cm^{-2}\,H^{-1}}$)
      only constrains the total volume $V_{\rm tot}$ 
      of this grain component:
      $A(\lambda)/\NH 
       = 1.086\,\NH^{-1} \int C_{\rm ext}(a,\lambda) 
         \left(dn/da\right)da
       = 1.086\,\NH^{-1} V_{\rm tot} \left(C_{\rm ext}/V\right)$,
      where $\NH$ is the hydrogen column density. 
      In contrast, the near and mid IR intensity $I_\lambda$ 
      is sensitive to the grain heat capacity ($\propto$\,$a^3$) 
      which determines the maximum temperature to which 
      the grain can reach when illuminated by a radiation field:
      $I_\lambda = \NH^{-1} \int C_{\rm abs}(a,\lambda) 
      \left(dn/da\right)da \int B_\lambda(T[a])\left(dP/dT\right)dT$
      where $C_{\rm abs}(a,\lambda)$ is the absorption cross section 
      for a grain of radius $a$ at wavelength $\lambda$
      (for grains in the Rayleigh limit, 
       $C_{\rm abs}$\,$\approx$\,$C_{\rm ext}$),        
      $B_\lambda$ is the Planck function,
      $dP/dT$ is the dust temperature distribution function.
      Evidently, $dP/dT$ is a sensitive function of
      grain size $a$ (see Draine \& Li 2001, Li \& Draine 2001b;
      and \S3, Fig.\,6 of this paper). 
\end{itemize}

In the following sections of this review, we will focus on
the ultrasmall grain component ($a$$\ltsim$$25\nm$), 
with particular emphasis on the photophysical processes 
including the stochastic heating and the vibrational excitation
of PAH molecules, and the excitation of the photoluminescence of 
silicon nanoparticles. In astrophysical literature, 
one frequently encounters terms like ``ultrasmall grains'', 
``very small grains'', ``large molecules'', ``tiny grains'', 
and ``nanoparticles''; to avoid confusion, we emphasize here 
that they are synonymous. In the following we will use the term
``nanoparticles'' -- by ``nanoparticles'' we mean grains
of a few angstroms to a few tens of nanometers in radius.

In \S2 we summarize the direct and indirect evidence
for the existence of nanoparticles in the ISM. 
The vibrational excitation of nanoparticles 
is detailed in \S3. In \S4 we discuss the physics 
regarding the excitation of photoluminescence.
In \S5 we present an overview of the nanoparticle
species known or proposed to exist in interstellar space.
Concluding remarks are given in \S6.

\vspace{-3.0mm}   
\section{Nanoparticles in Interstellar Space}
\vspace{-2.0mm}
The discovery of presolar nanodiamonds (see \S5.4)
and TiC nanocrystals (see \S5.5) in primitive meteorites 
implies that there must exist such nano-sized species
in interstellar space.  
In addition, as already mentioned in \S1, 
the presence of nanoparticles
in the ISM is clearly indicated by ---
\begin{itemize}
\vspace*{-0.3mm}
\item \underline{(1) The ubiquitous distinctive set of 
      ``UIR'' emission bands at 3.3, 6.2, 7.7,}\\ 
      \underline{8.6, and 11.3$\mum$}. 
      This emission, accounting for $\simali$20\% of 
      the total power radiated by dust, is closely explained
      by transiently heated PAHs with an abundance 
      ${\rm C/H}$\,$\approx$\,45
      parts per million (ppm) and a log-normal size distribution 
      peaking at $\simali$6$\Angstrom$ (with $\simali$100 carbon atoms;
      see Li \& Draine 2001b for details).
\vspace*{-0.5em}
\item \underline{(2) The mid-IR emission at $\lambda \ltsim 60\mum$.}
      This emission, first discovered by the IRAS broadband 
      photometry at 12 and 25$\mum$ and later confirmed by 
      the COBE-DIRBE observations, cannot be explained by
      large grains ($a\gtsim$$250\Angstrom$) heated by 
      the interstellar radiation field to equilibrium temperatures 
      $\approx$\,15--25$\K$ since the predicted emission intensities
      from large grains are smaller than the IRAS 12 and 25$\mum$ 
      emission intensities by several orders of magnitude. 
      In the diffuse ISM, the emission
      at $\lambda$$\ltsim$$60\mum$ accounts for $\simali$35\% of 
      the total power radiated by dust. It is well recognized that
      this emission arises from nanoparticles 
      ($a\ltsim$$250$$\Angstrom$) stochastically heated by 
      single UV/visible photons to temperatures
      significantly higher than their time-averaged temperatures
      (
       see Draine \& Li 2001 and references therein).
       Even at 60$\mum$, the nanoparticle component contributes 
       $\simali$70\% of the total emission power of the diffuse ISM 
       detected by the COBE-DIRBE photometers (see Fig.\,8 of
       Li \& Draine 2001b).
\vspace*{-0.5mm}
\item \underline{(3) To a lesser degree, the far-UV extinction rise.}
      The far-UV interstellar extinction continues to rise up to 
      $\lambda = 0.1\mum$, and there does not appear to 
      be any evidence of saturation even at this wavelength
      (see Whittet 2003). Since it is generally true that a grain
      absorbs and scatters light most effectively at wavelengths 
      comparable to its size $\lambda \approx 2\pi a$ (Kr\"ugel 2003), 
      we can therefore conclude that there must be 
      appreciable numbers of interstellar grains with 
      $a$\,$\ltsim$\,$0.1\mum/2\pi$\,$\approx$\,$16\nm$
      (e.g., in the size distribution derived by 
       Weingartner \& Draine [2001a], grains smaller than 2$\nm$
       provides $\approx 80\%$ of the total surface area, although
       they contain only $\approx 6\%$ of the total dust mass). 
      However, as remarked earlier (see \S1), the far-UV extinction
      curve is not able to tell us the details of the size distribution
      of the nanoparticle component. 
\item \underline{(4) The ``anomalous'' Galactic foreground 
      microwave emission in the 10---}\\
      \underline{100$\GHz$ region.} 
      This emission, discovered unexpectedly in recent experiments 
      to study the angular structure in the cosmic background
      radiation, was found to be positively correlated with 
      interstellar dust, as traced by the 100$\mum$ IRAS map 
      or the 140$\mum$ COBE-DIRBE map 
      (see Draine 1999 and references therein). 
      However, the dust thermal emission
      at microwave frequencies extrapolated from the 100--3000$\mum$
      far-IR emission radiated by large grains in thermal equilibrium
      with the interstellar radiation field falls far below 
      the observed microwave emission (Draine 1999).
      This ``anomalous'' emission also significantly exceeds 
      the free-free emission from interstellar plasma 
      (Draine \& Lazarian 1998a). As described by Draine and his
      coworkers (Draine \& Lazarian 1998b; Draine \& Li 2003),
      a number of physical processes, including collisions with 
      neutral atoms and ions, ``plasma drag'' (due to interaction 
      of the electric dipole moment of the grain with the electric 
      field produced by passing ions), and absorption and emission 
      of photons, can drive nanoparticles to rapidly rotate, with rotation
      rates reaching tens of GHz. The rotational electric dipole emission 
      from these spinning nanoparticles, the very same grain component
      required to account for the ``UIR'' emission and the IRAS 
      12 and 25$\mum$ emission, was shown to be capable of
      accounting for the ``anomalous'' microwave emission
      (Draine \& Lazarian 1998a,b; Draine 1999).
\item \underline{(5) The photoelectric heating of the diffuse ISM.} 
      Grains are long thought to be an important energy 
      source for the interstellar gas through ejection of 
      photoelectrons 
      since (a) photons with energies below the ionization potential 
      of H ($\simali$13.6$\eV$) do not couple directly to the gas; 
      and (b) other heating sources such as cosmic rays, 
      magnetic fields, and turbulence are unimportant as 
      a global heating source for the diffuse ISM.

      The photoelectric heating starts from the absorption
      of a far-UV photon by a dust grain, followed by
      ejection of an electron which then collisionally 
      heats the interstellar gas by transferring (to the gas) 
      the excess energy left over after overcoming the work function 
      (the binding energy of the electron to the grain) 
      and the electrostatic potential of the grain (if it is charged).

      In the diffuse ISM, nanoparticles (and in particular, 
      angstrom-sized PAH molecules) are much more efficient 
      in heating the gas than large grains 
      (see Tielens \& Peeters 2002 and references therein)
      since (a) the mean free path of an electron in a solid 
      is just $\simali$10$\Angstrom$ and therefore photoelectrons
      created inside a large grain rarely reach the grain surface;
      and (b) the total far-UV absorption is dominated by 
      the nanoparticle component.
      Recent studies show that grains smaller than $10\nm$
      are responsible for $\gtsim 96\%$ of the total photoelectric 
      heating of the gas, with half of this provided by
      grains smaller than 15$\Angstrom$
      (Bakes \& Tielens 1994; Weingartner \& Draine 2001b). 
\item \underline{(6) The Extended Red Emission (ERE) ?}
      This emission, first detected in the Red Rectangle 
      (Schmidt, Cohen, \& Margon 1980), is characterized 
      by a broad, featureless band between 
      $\simali$5400$\Angstrom$ and 9500$\Angstrom$,
      with a width 
      $600\Angstrom \ltsim {\rm FWHM} \ltsim 1000\Angstrom$
      and a peak of maximum emission at
      $6100\Angstrom \ltsim \lambda_{\rm p} \ltsim 8200\Angstrom$, 
      depending on the physical conditions of the environment 
      where the ERE is produced. 
      The ERE has been seen in a wide variety of dusty environments: 
      the diffuse ISM of our Galaxy, reflection nebulae, 
      planetary nebulae, HII regions, and other galaxies 
      (see Witt \& Vijh 2004 for a review). 
      The ERE is generally attributed to photoluminescence (PL) 
      by some component of interstellar dust,
      powered by UV/visible photons.

      The observational evidence shows that 
      (a) the ERE carriers must have a photon conversion 
      efficiency $\etapl$ (the number ratio of PL photons 
      to exciting photons) substantially larger than 10\% 
      as estimated from the correlation of ERE intensity 
      with HI column density at high Galactic latitudes
      (Gordon, Witt, \& Friedmann 1998),
      and (b) the carriers can be easily modified or destroyed 
      by intense UV radiation (Witt 2000).
      This suggests that the ERE carriers are very likely
      in the nanometer size range because (a) in general,
      nanoparticles are expected to luminesce efficiently 
      through the recombination of the electron-hole pair
      created upon absorption of an energetic photon,
      since in such small systems the excited electron 
      is spatially confined and the radiationless transitions 
      that are facilitated by Auger and defect related recombination 
      are reduced (see \S4);
      and (b) small nanoparticles may be photolytically 
      more unstable and/or more readily photoionized in 
      regions where the radiation intensity exceeds certain 
      levels of intensity and hardness, and thus resulting in 
      both a decrease in the ERE intensity and a redshift of 
      the ERE peak wavelength,
      since (i) photoionization would 
      quench the luminescence of nanoparticles,
      and (ii) the smaller grains would be selectively 
      removed due to size-dependent photofragmentation
      (Smith \& Witt 2002).

      Proposed ERE carriers include 
      (a) ``classic'' submicron-sized carbonaceous materials: 
      HAC (Duley 1985; Witt \& Schild 1988), 
      QCC (Sakata et al.\ 1992), and coal (Papoular et al.\ 1996); 
      (b) nanometer-sized carbon-based materials: 
      PAHs (d'Hendecourt et al.\ 1986), 
      and carbon nanoparticles (Seahra \& Duley 1999);
      (c) nanometer-sized silicon-based materials:
      crystalline silicon nanoparticles (Witt et al.\ 1998;
      Ledoux et al.\ 1998, 2001; Smith \& Witt 2002);
      and (d) particle-bombarded silicate grains
      (Koike et al.\ 2002).

      The carbon-based models appear to be ruled out:
      (a) submicron-sized carbon materials appear to 
      be unable to simultaneously match the observed 
      ERE spectra and the required PL efficiency (Witt et al.\ 1998);
      (b) although high photoluminescence efficiencies can be 
      obtained by PAHs, the lack of spatial correlation 
      between the ERE and the PAH IR emission bands 
      in some regions (see Li \& Draine 2002a and references therein),
      together with nondetection of ERE emission in reflection nebulae 
      illuminated by stars with effective temperatures
      $T_{\star}$$<$$7000\K$ (Darbon, Perrin, \& Sivan 1999), 
      whereas PAHs emission bands have been seen
      in such regions (e.g., see Uchida, Sellgren, \& Werner 1998)
      and are expected for the PAH emission model 
      (Li \& Draine 2002b),
      seems to argue against PAHs as ERE carriers;
      (c) the carbon nano-cluster hypothesis put forward 
      by Seahra \& Duley (1999) appears to be invalid as 
      indicated by nondetection in NGC 7023 of 
      the predicted 1$\um$ ERE peak (Gordon et al.\ 2000),
      although they argued that these carbon nanoparticles 
      with mixed $sp^2/sp^3$ bonding would be able to 
      meet both the ERE profile and the PL efficiency requirements.

      The silicon nanoparticle (SNP) model, originally  
      proposed by Witt et al.\ (1998) and Ledoux et al.\ (1998),
      seems promising. Experimental data show that SNPs
      provide so far the best match to the observed ERE spectra 
      and to the quantum efficiency requirement.
      However, this model also has difficulty:
      we calculated the thermal emission 
      expected from such particles, 
      both in a reflection nebula such as NGC 2023 
      and in the diffuse ISM;
      we found that SNPs would produce a strong emission 
      feature at 20$\mum$ which is not seen in the observational
      spectra; therefore we concluded that 
      if the ERE is due to SNPs, 
      they must be either in clusters 
      or attached to larger grains 
      (see Li \& Draine 2002a for details).
\end{itemize}

\vspace*{-1.6em}
\section{Excitation of Vibrational Transitions}
\vspace*{-0.4em}
For isolated, free-flying nanoparticles or large molecules 
in many astrophysical environments, photon absorption is 
often the dominant excitation process. 
Prior to the absorption of an energetic photon,
a molecule stays at the ground electronic state, 
which is $S_0$ (the lowest singlet state) for neutrals
and $D_0$ (the lowest doublet state) for ions,
containing very little vibrational energy. 
After photoabsorption occurs, the molecule is excited
to an upper electronic state $S_n$ for neutrals 
or $D_n$ for ions ($n$$>$1). The electronically excited molecule
has 3 major competing decay channels to relax its energy: 
radiation, photoionization, and photodissociation.
In this section we will focus on the radiation process
and use PAHs as an example.

\subsection{An Overview of the Photoexcitation and Emission Processes}
Once the molecule is electronically excited, both
radiative and radiationless transitions will occur
between its vibrational modes, electronic and 
vibrational states (see Birks 1970 for detailed
discussions on neutral molecules):
\begin{itemize}
\item the molecule will undergo 
      \underline{internal vibrational redistribution (IVR)},
      an iso-\\energetic radiationless process between 
      vibrational modes, 
      which rapidly ($\simali$$10^{-12}$--$10^{-10}\s$)
      spreads the absorbed energy among its vibrational 
      degrees of freedom;
\vspace{-1mm}
\item the molecule will undergo 
      \underline{internal conversion (IC)},
      a radiationless transition between 
      electronic states of the {\it same} multiplicity, 
      which very quickly ($\simali$$10^{-12}$--$10^{-8}\s$)
      transfers the electronic energy to a highly vibrationally 
      excited state of the lower lying electronic state
      (i.e., $S_i$$\rightarrow$$S_j$ for neutrals,
       and $D_i$$\rightarrow$$D_j$ for ions where $i$$>$$j$);
\vspace{-1mm}
\item the molecule will also undergo 
      \underline{intersystem crossing (ISC)},
      a rapid radiationless transition 
      between electronic states of {\it different} multiplicity 
      (i.e., $S_i$$\rightarrow$$T_j$ for neutrals,
       and $D_i$$\rightarrow$$Q_j$ for ions where $i$$>$$j$);
\vspace{-1mm}
\item the molecule will undergo \underline{fluorescence},
      a radiative electronic transition ($\simali$$10^{-7}\s$)
      between states of the {\it same} multiplicity, 
      which results in the emission of a visible photon
      (i.e., $S_i$$\rightarrow$$S_j$ for neutrals,
       and $D_i$$\rightarrow$$D_j$ for ions where $i$$>$$j$;
       for many molecules, this transition mainly takes place
       between the first excited electronic state $S_1$ or $D_1$
       and the ground state $S_0$ or $D_0$, as a consequence of
       rapid internal conversion which results in a practically
       complete conversion of the initial excitation energy
       to vibrationally excited levels of the $S_1$ or $D_1$ 
       and $S_0$ or $D_0$ electronic states);
\vspace{-1mm}
\item the molecule will also undergo \underline{phosphorescence}
      with emission of a visible photon,
      a very slow ($\simali$a few seconds) 
      radiative electronic transition between 
      states of {\it different} multiplicity
      (i.e., $T_i$$\rightarrow$$S_j$ for neutrals where $i$$>$$j$;
       for ions, this transition [i.e. $Q_i$$\rightarrow$$D_j$] 
       is not important 
       since internal conversion dominates over intersystem crossing 
       so that the molecule ends up in highly vibrationally excited 
       levels of the ground electronic state $D_0$ [Leach 1987]);
\vspace{-1mm}
\item finally, the molecule will undergo \underline{IR emission},
      a radiative vibrational transition ($\simali$$0.1\s$)
      between higher and lower vibrational states 
      of the same electronic state.
      This transition mainly takes place in the ground electronic
      state $S_0$ or $D_0$ because of rapid internal conversion
      and electronic fluorescence which ultimately drive the molecule 
      down to the ground electronic state with a high vibrational energy,
      except for some neutral molecules (e.g. chrysene C$_{18}$H$_{12}$)
      a large fraction of this transition occurs in the first excited 
      triplet state $T_1$ as a result of rapid intersystem crossing.
      Note that in Figure 1 we only plot the $\Delta$v=1 vibrational
      transitions (where v is the vibrational quantum number)
      given by the harmonic oscillator selection rule;
      but overtone transitions with $\Delta$v=2,3,... are also
      allowed when account is taken of anharmonicity. 
      We should stress that in astrophysical literature,
      vibrational transitions are sometimes also called 
      ``vibrational fluorescence'' (e.g. see Allamandola et al.\ 1989); 
      in this case, we shall call the fluorescence process
      described above ``electronic fluorescence''. 
\end{itemize}
In addition, for highly isolated molecules, there could exist 
another two radiative energy decay channels -- 
the \underline{inverse fluorescence} (Leach 1987) 
and the \underline{recurrent fluorescence} 
(L\'eger, Boissel, \& d'Hendecourt 1988).
The former results from a transition from 
a high vibrational level of a lower electronic state 
to a higher electronic state.  
The latter, also known as the \underline{Poincar\'e fluorescence},
was postulated by L\'eger et al.\ (1988) as resulting from 
an inverse electronic conversion 
(i.e., the partial conversion of the vibrational energy
of the ground electronic state into electronic excitation),
followed by emitting a visible photon just like the ordinary
(electronic) fluorescence process. 
However, the Poincar\'e fluorescence
may have a quantum yield larger than one 
(i.e., several fluorescence photons 
can be emitted during one photoabsorption event 
since the molecule can oscillate many times between 
the electronic ground state and the excited state 
provided the absorbed energy is high enough; 
see L\'eger et al.\ 1988). 

\vspace{2mm}
Figures 1 and 2 give a schematic overview of the above-described 
main processes following the absorption of an energetic photon 
respectively for a neutral molecule and a molecular cation.

\vspace{2mm}
The stochastic heating of ultrasmall grains
has been extensively studied in literature
(see Draine \& Li 2001 and references therein).
All studies prior to the identification of PAH
molecules as the carrier of the ``UIR'' bands
used the ``thermal'' approach (i.e., the vibrationally 
excited grain is described as a system having an internal 
temperature; see \S3.3). The issue of ``thermal''
versus ``non-thermal'' arose when L\'eger and his coworkers 
used the ``thermal'' approach to model the vibrational
excitation of PAHs to explain the ``UIR'' bands 
(L\'eger \& Puget 1984; L\'eger, d'Hendecourt, \& D\'efourneau 1989),
while Allamandola, Tielens, \& Barker (1985, 1989) 
took a statistical approach
and questioned the validity of the ``thermal'' approximation.
Barker \& Cherchneff (1989), d'Hendecourt et al.\ (1989),
Schutte, Tielens, \& Allamandola (1993), and Cook \& Saykally (1998)
found that the thermal approximation was valid for computing 
thermal emission spectra.
Below we will discuss the recent efforts we took to 
model the PAH excitation and de-excitation processes
(Draine \& Li 2001) using both the ``exact-statistical'' 
approach (\S 3.2) and the ``thermal-approximation'' 
approach (\S3.3).

\begin{figure} 
\vspace*{-1.2em}
\hspace*{-0.8em}
\centerline{
\psfig{figure=neuexc.cps,height=16cm,width=14cm}}
\vspace*{-1.5em}
\caption{
        \normalsize
        Step ladder model of a large neutral molecule.
        The schematic energy levels involve
        an electronic part (singlets $S_0$, $S_1$, $S_2$ ...
        and triplets $T_1$, $T_2$ ...)
        and a vibrational part (v$_1$, v$_2$, v$_3$...).
        Upon absorption of an UV/visible photon,
        the molecule makes an electronic transition from 
        the ground state $S_0$ to an upper state $S_n$, 
        followed by
        (1) radiationless internal vibrational redistribution,
        (2) radiationless electronic transitions 
        (internal conversion [IC] and intersystem crossing [ISC]),
        (3) radiative electronic transitions 
        (fluorescence and phosphorescence),
        (4) radiative inverse electronic transitions
        (inverse fluorescence, and Poincar\'e fluorescence
        resulted from the inverse internal conversion [IIC]),
        and (5) IR vibrational transitions.
        }
\end{figure}

\begin{figure}[] 
\vspace*{-1.2em}
\hspace*{-0.8em}
\centerline{
\psfig{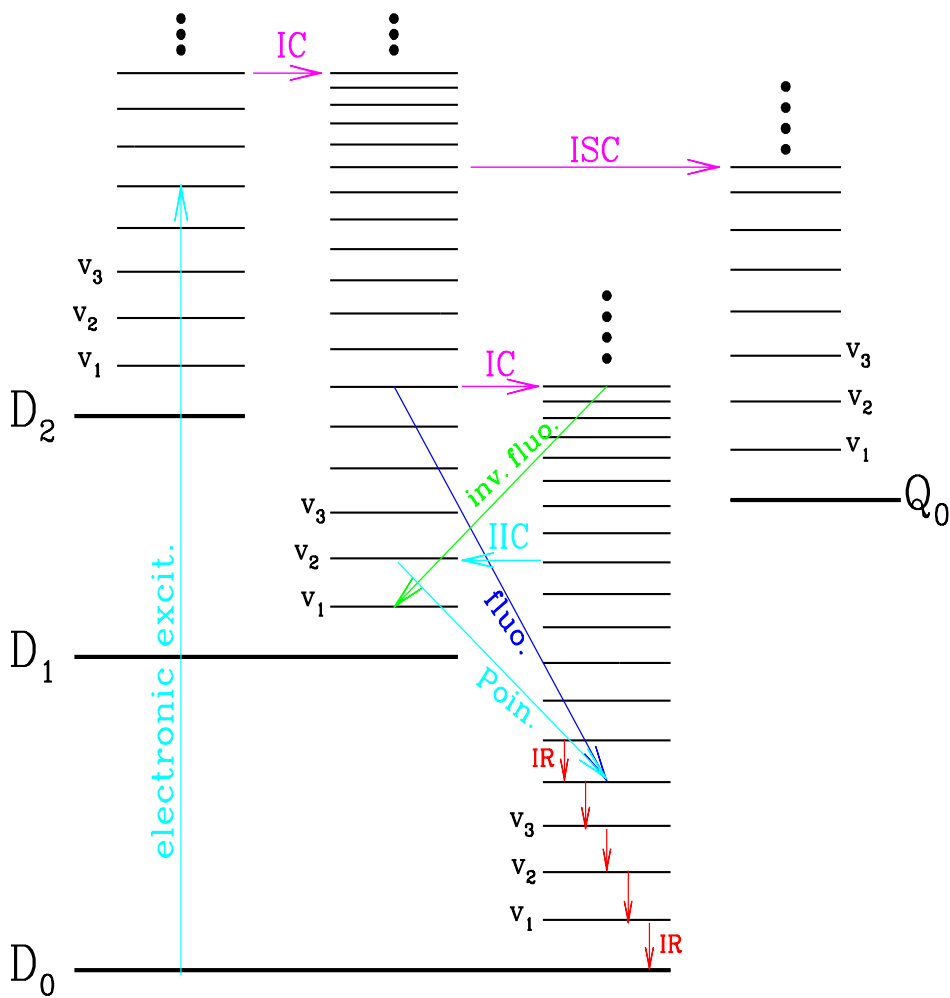}}
\vspace*{-1.5em}
\caption{
        \normalsize
        Same as Figure 1 but for a large molecular cation
        (with one unpaired electron).
        The schematic energy levels involve
        an electronic part (doublets $D_0$, $D_1$, $D_2$ ...
        and quartets $Q_0$, $Q_1$, $Q_2$ ...)
        and a vibrational part (v$_1$, v$_2$, v$_3$...).
        In comparison with neutral molecules,
        radiative electronic transitions (fluorescence
        and phosphorescence) for ions are not important
        since internal conversion quickly transfers 
        almost all of the initial excitation energy
        to the vibrationally excited ground electronic state $D_0$
        (see Leach 1987; Allamandola et al.\ 1989).
        Therefore, IR emission becomes almost the only deactivation route.
        }
\end{figure}

\subsection{``Exact-Statistical'' Approach}
\vspace*{-0.5em}
As discussed in \S3.1, in most cases soon after 
the photoabsorption an isolated nanoparticle 
(or large molecule) converts almost all of 
the initial photoexcitation energy to vibrational energy of 
the highly vibrationally excited ground electronic state, 
and hence for both neutrals and ions, IR emission is always 
the dominant deactivation process.
Therefore, it is reasonable to model the stochastic heating
of a nanoparticle in terms of pure vibrational transitions.

Ideally, if both the vibrational energy levels and the level-to-level
transition probabilities were known, we could (at least in principle)
solve for the statistical steady-state populations of the different
energy levels of grains illuminated by a known radiation field.
However, this level of detailed information is generally unavailable,
for even the smallest and simplest PAH molecules.

\begin{figure}[]  
\vspace*{-0.5em}
\hspace*{-1.5em}
\centerline{
\psfig{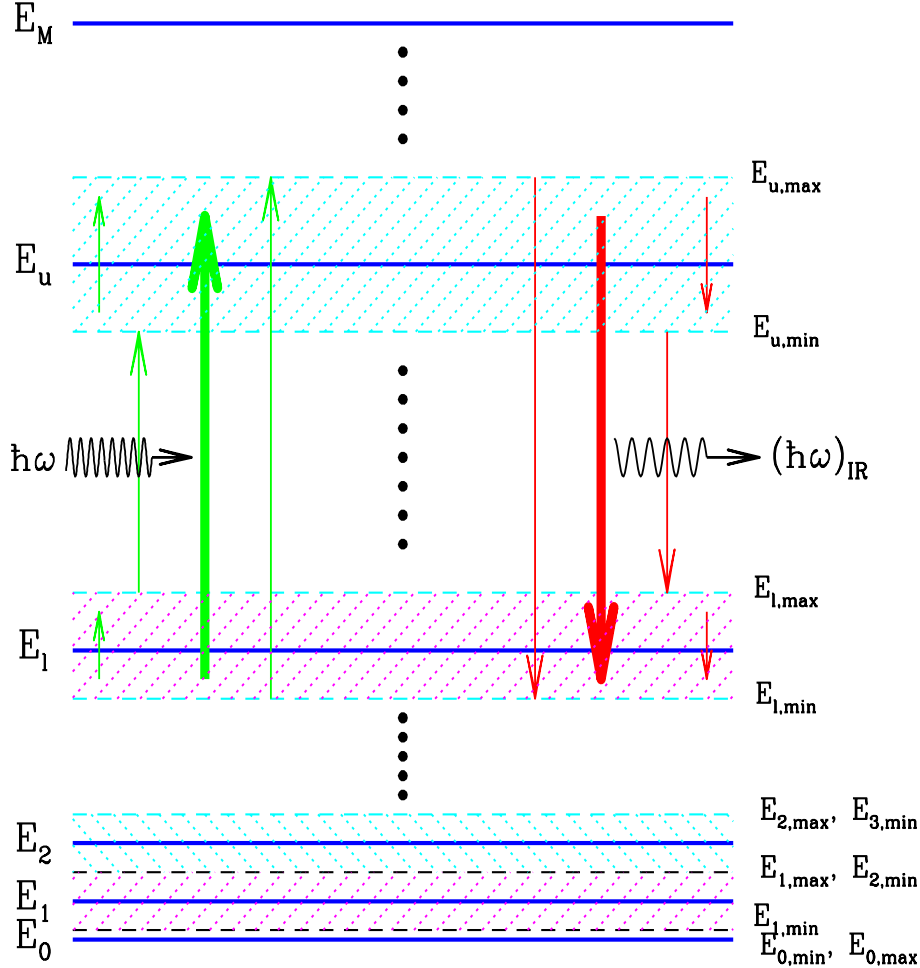}}
\vspace*{-2.9em}
\caption{
        \label{fig:dlexc}
        \normalsize
        Schematic diagram of the vibrational excitation
        and relaxation processes in nanoparticles (or large
        molecules). The vibrational energy levels are
        grouped into $\left(\Nb+1\right)$ ``bins'' $j=0,...,\Nb$, 
        where the $j$-th bin is $[E_{j,\min},E_{j,\max})$, 
        with representative energy $E_j\equiv(E_{j,\min}+E_{j,\max})/2$, 
        and width $\Delta E_j\equiv \left(E_{j,\max}-E_{j,\min}\right)$.
        The ground state is set at $E_{0,\min}=E_{0,\max}=E_0=0$
        (see Appendix B of Draine \& Li 2001 for the procedure 
         for specifying the bins).
        The wiggly arrow represents an event of photon
        absorption ({\it left}) or emission ({\it right}).
        The small arrows within the bin $u$ and bin $l$
        represent the ``intrabin'' transitions.
        With the bin-to-bin transition rates determined,
        we can solve for the statistical steady-state populations 
        of the different energy states of grains illuminated by 
        a known radiation field,
        and then calculate the resulting IR emission spectra.
        }
\end{figure}

Draine \& Li (2001) developed an ``exact-statistical'' theory
for modeling the photoexcitation and emission processes of 
nanoparticles. In this theory, the state of the grain is 
characterized by its vibrational energy $E$.
Since there are too many vibrational energy levels 
to consider individually, they are grouped into 
$\left(\Nb+1\right)$ ``bins'' $j=0,...,\Nb$, 
where the $j$-th bin is $[E_{j,\min},E_{j,\max})$, 
with representative energy $E_j\equiv(E_{j,\min}+E_{j,\max})/2$, 
and width $\Delta E_j\equiv \left(E_{j,\max}-E_{j,\min}\right)$
(see Fig.\,3 for illustration). 
Let $P_j$ be the probability of finding the grain in bin $j$.
The probability vector $P_j$ evolves according to
$dP_i/dt = \sum_{j\neq i} \bT_{ij} P_j - \sum_{j\neq i} \bT_{ji}P_i$
for $i=0,1,...,\Nb$,
where the transition matrix element $\bT_{ij}$ is
the probability per unit time for
a grain in bin $j$ to make a transition to one of the levels 
in bin $i$. If we define the diagonal elements of $\bT$ to be
$\bT_{ii} \equiv -\sum_{j\neq i}\bT_{ji}$,
then under the steady state condition 
(i.e. $dP_i/dt = 0$ for $i=0,1,...,\Nb$) 
the state probability evolution equation becomes
$\sum_{j=0}^{\Nb} \bT_{ij}P_j = 0$ for $i=0,...,\Nb$.
Combining this with the normalization condition 
$\sum_{j=0}^{\Nb} P_j=1$, we obtain a set of $\Nb$ linear equations 
for the first $\Nb$ elements of $P_j$:
$\sum_{j=0}^{\Nb-1}\left(\bT_{ij}-\bT_{i\Nb}\right)P_j = -\bT_{i\Nb}$
for $i=0,...,\Nb-1$, which we solve using the bi-conjugate gradient 
(BiCG) method. The remaining undetermined element $P_{\Nb}$ 
is obtained by
$P_{\Nb} = -\left(\bT_{\Nb\Nb}\right)^{-1} 
\sum_{j=0}^{\Nb-1} \bT_{\Nb j}P_j$.

For a given starlight energy density $u_E$,
the state-to-state transition rates $\bT_{ji}$ for transitions
$i$$\rightarrow$$j$ can be determined from photon absorptions 
and photon emissions.
The rate for upward transitions $l$$\rightarrow$$u$ 
is just the absorption rate of photons with such an energy
that they excite the grain from bin $l$ {\it just} to bin $u$.
If the bin width is {\it sufficiently small}
(i.e., if $\max\left[\Delta E_l, \Delta E_u\right] 
\ll \left[E_u-E_l\right]$),
the $l$$\rightarrow$$u$ excitation rate is simply 
$\bT_{ul} \approx C_{\abs}(E)\,c\,u_E \Delta E_u/\left(E_u-E_l\right)$
for $u$$<$$\Nb$, where $C_\abs(E)$ is the grain absorption 
cross section at wavelength $\lambda$=$hc/E$ 
($h$ is the Planck function and $c$ is the speed of light); and
$\bT_{Ml} \approx C_{\abs}(E)\,c\,u_E \Delta E_M/\left(E_M-E_l\right)
+ \int_{E_M-E_l}^\infty dE\,C_{\abs}(E)\,c\,u_E/\left(E_M-E_l\right)$,
where the integral takes energy absorbed in transitions to levels 
beyond the highest bin and allocates it to the highest bin ($M$).
For the special case of transitions 
$u-1$$\rightarrow$$u$
we include ``intrabin'' absorptions:
$\int_0^{\Delta E_{u\!-\!1}}\,dE\, 
\left(1-\frac{E}{\Delta E_{u\!-\!1}}\right) 
C_{\abs}(E)\,c\,u_E/\left(E_u\!-\!E_{u\!-\!1}\right)$.
Correction for finite bin width, which is important when 
the treatment is applied to grains with radii $a$$\gtsim$$50\Angstrom$, 
has been made by Draine \& Li (2001) by introducing a $G_{ul}(E)$
factor (see Eqs.[15-25] of Draine \& Li 2001).

The rates for downward transitions $u$$\rightarrow$$l$
can be determined from a detailed balance analysis
of the Einstein $A$ coefficient.
Similarly, if the bin width is sufficiently small,
the $u$$\rightarrow$$l$ de-excitation rate can be approximated as
$\bT_{lu} \approx 
\frac{8\pi}{h^3c^2} \frac{g_l}{g_u}
\frac{\Delta E_u}{E_u-E_l} 
E^3$$\times$$C_{\abs}(E) \left[1+\frac{h^3c^3}{8\pi E^3}u_E\right]$
for $l$$<$$u-$1, where the $u_E$-containing term is the contribution 
of stimulated emission, and the degeneracies $g_u$ and $g_l$ 
are the numbers of energy states in bins $u$ and $l$, respectively:
$g_j\equiv N(E_{j,\max})-N(E_{j,\min})
\approx \left(\frac{dN}{dE}\right)_{E_j}
\Delta E_j$, where $\left(\frac{dN}{dE}\right)_{E_j}$
is the vibrational density of states at internal energy $E_j$, 
which corresponds to the number of ways of distributing this energy 
between different modes of this grain.
Again, we refer the reader to Draine \& Li (2001) 
for finite bin width corrections as well as 
``intrabin'' radiation consideration 
(see Eqs.[29-31] of Draine \& Li 2001).

It is seen from the above discussions that 
we require only $C_{\abs}(E)$, the degeneracies $g_j$, 
and the starlight spectrum $u_E$ to completely 
determine the transition matrix $\bT_{ij}$.
A molecule containing $N_a$ atoms will have 
$N_m=3N_a-6$ distinct vibrational modes
(plus 3 translational degrees of freedom
and 3 rotational degrees of freedom).
If the molecule is approximated as a set of 
$N_m$ harmonic oscillators, and the frequencies of 
all normal modes of this molecule are known,
we can calculate $N(E)$, the number of distinct vibrational states 
with total vibrational energy less than or equal to $E$, 
using the Beyer-Swinehart algorithm (Beyer \& Swinehart 1973;
Stein \& Rabinovitch 1973). So far, the frequencies of 
these normal modes have been computed only for a small number 
of PAHs, with some frequencies determined experimentally, 
but mode spectra are not yet available for most PAHs of interest.

Since exact densities of states are often unknown for
interstellar PAHs, the Whitten \& Rabinovitch (1963)
approximation, a semi-empirical expression, 
has been extensively used in literature:
$\rho(E) = \left(E+\xi\,E_z\right)^{N_m-1}$/
$\left[(N_m-1)!\prod^{N_m}_{i=1}h\nu_i\right]$
where $\rho(E)$ is the density of states
(the number of accessible vibrational states per
unit energy) at internal energy $E$,
$E_z=\sum^{N_m}_{i=1}\left(h\nu_i/2\right)$ 
is the total zero point energy of the molecule,
$\nu_i$ is the vibrational frequency,
and 0$<$$\xi$$<$1 is an empirical correction factor. 

In contrast, the Draine \& Li (2001) ``exact-statistical'' 
theory does not need this approximation; instead, they
calculate the ``theoretical'' mode spectrum from the Debye
model. A PAH molecule containing $N_\C$ C atoms and
$N_\rmH$ H atoms is treated by Draine \& Li (2001) 
as having 5 different types of vibration: 
(1) ($N_{\rm C}-2$) out-of-plane C--C modes at 
$\lambda_{\C\C,{\rm op}}^{-1} = \kB\Theta_{\rm op}/hc
\approx \left(16.7\mum\right)^{-1}\approx 600\cm^{-1}$
given by a two-dimensional Debye model with a Debye temperature
$\Theta_{\rm op}\approx950\K$,
where $\kB$ is the Boltzmann constant,
(2) $2\,(N_\C-2)$ in-plane C--C modes at
$\lambda_{\C\C,{\rm ip}}^{-1} = \kB\Theta_{\rm ip}/hc
\approx \left(5.7\mum\right)^{-1}\approx 1740\cm^{-1}$
given by a two-dimensional Debye model 
with a Debye temperature $\Theta_{\rm ip}\approx2500\K$,
(3) $N_\rmH$ out-of-plane C--H bending modes at 
$\lambda_{\C\rmH,{\rm op}}^{-1}=(11.3\mum)^{-1}\approx 886\cm^{-1}$,
(4) $N_\rmH$ in-plane C--H bending modes at 
$\lambda_{\C\rmH,{\rm ip}}^{-1}=(8.6\mum)^{-1}\approx 1161\cm^{-1}$,
and (5) $N_\rmH$ C--H stretching modes at 
$\lambda_{\C\rmH,{\rm str}}^{-1}=(3.3\mum)^{-1}\approx 3030\cm^{-1}$.
The ``synthetic'' mode spectrum for C$_{24}$H$_{12}$ is in 
excellent agreement with the actual mode spectrum of coronene
(see Fig.\,1 of Draine \& Li 2001).
Similarly, a silicate grain containing $N_a$ atoms is treated as  
having $2\,(N_a-2)$ vibrational modes distributed according to 
a two-dimensional Debye model with
a Debye temperature $\Theta=500\K$,
and $(N_a-2)$ modes described by 
a three-dimensional Debye model with $\Theta=1500\K$.

From the ``synthetic'' model mode spectrum we can obtain
the vibrational density of states and hence the ``degeneracy'' 
$g_j$, the number of distinct quantum states included in bin $j$.
We note that the densities of states computed for C$_{24}$H$_{12}$, 
using both the actual normal mode spectrum for coronene 
and the model normal mode spectrum for C$_{24}$H$_{12}$ 
(see Fig.\,1 of Draine \& Li 2001) are essentially identical 
for $E/hc\gtsim 300\cm^{-1}$ (see Fig.\,4 of Draine \& Li 2001).
With $g_j$ derived from the model mode spectrum, 
and the $j$$\rightarrow$$i$ ($j$$<$$i$) 
excitation rates $\bT_{ij}$ calculated from 
a known radiation field with energy density $u_E$, 
we can determine the $i$$\rightarrow$$j$ ($i$$>$$j$)
de-excitation transition rates $\bT_{ji}$.
Solving the steady-state state probability evolution equation
$\sum_{j\neq i} \bT_{ij} P_j = \sum_{j\neq i} \bT_{ji}P_i$
for $i=0,1,...,\Nb$, we are able to obtain the steady-state 
energy probability distribution $P_j$ 
and calculate the resulting IR emission spectrum.

\begin{figure}[] 
\vspace*{-1.0em}
\centerline{
\psfig{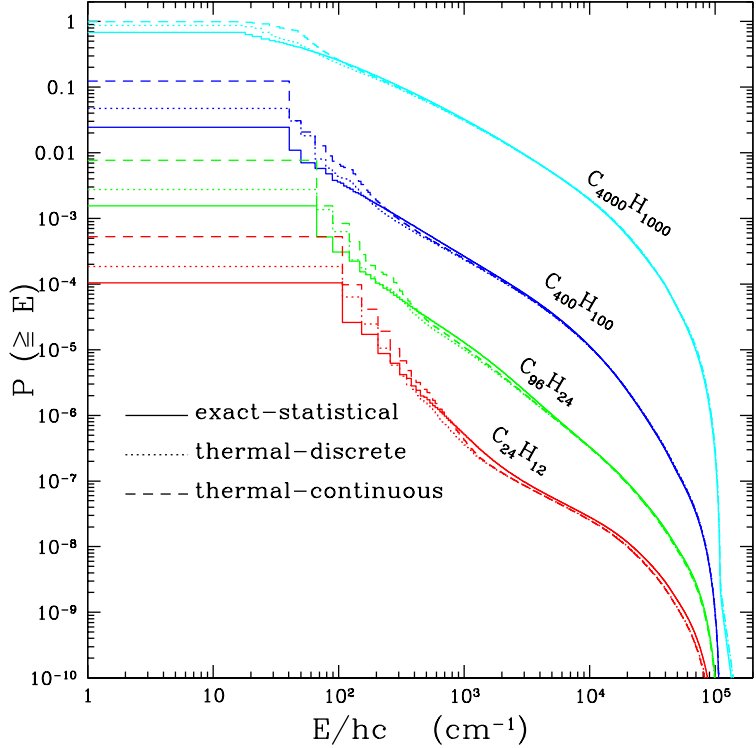}}
\vspace*{-1.0em}
\caption{
        \footnotesize
	The cumulative energy probability distributions for PAHs
        illuminated by the general ISRF
	computed using the {\it exact-statistical} model,
	the {\it thermal-discrete} model, 
	and the {\it thermal-continuous} model.
        Note that the lowest energy state ($E=0$), not shown here,
        has $P(E\geq 0)=1$. Taken from Draine \& Li (2001).
        }
\vspace*{-1.5em}
\end{figure}

In Figure 4 we present the {\it cumulative} energy probability 
distributions for selected PAHs excited by 
the general solar neighbourhood
interstellar starlight radiation field (ISRF) of
Mathis, Mezger \& Panagia (1983, hereafter MMP)
obtained from the ``exact-statistical'' model.
It is seen in Figure 4 that the probability of being in 
the ground state is very large for small grains:
for example, for the MMP radiation field, 
grains with $N_\C\ltsim4000$ spend most of their time at $E=0$.
The sharp drop at 13.6\,eV ($E/hc \approx 1.1\times 10^5$\,cm$^{-1}$)
is due to the radiation field cutoff at 912\,\AA\ and 
to the fact that multiphoton events are rare. 
The resulting IR emission spectra are displayed in 
Figure 5. The sawtooth features seen at long wavelengths
are due to our treatment of transitions from the lower excited 
energy bins to the ground state and first few excited states
(see \S5.1 and Appendix B of Draine \& Li 2001).

\begin{figure}[] 
\vspace*{-0.2em}
\centerline{
\psfig{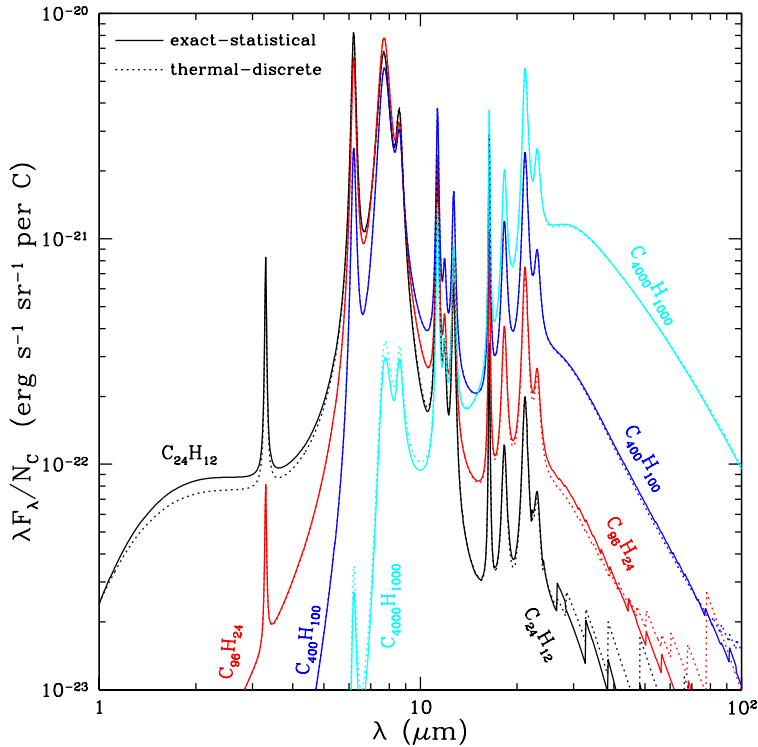}}
\vspace*{-0.8em}
\caption{
        \footnotesize
        \label{fig:IR_stat_exact}
        IR emissivities (per C atom) for selected ionized PAHs
	in the general ISRF
        calculated using the {\it exact-statistical} 
        and {\it thermal-discrete} models.
        Taken from Draine \& Li (2001).
	}
\vspace*{-1.0em}
\end{figure}

\subsection{``Thermal'' Approach}
\vspace*{-0.5em}
Given (1) the Debye model assumption 
about the vibrational mode spectrum, 
(2) the assumption that the absorption cross section 
$C_\abs(E)$ depends only on the photon energy, 
and (3) the assumption of rapid internal vibrational redistribution,
the ``statistical'' approach described in \S3.2 is ``exact''.
It is often desirable, however, to use an alternative ``thermal''
approximation which is less computationally demanding,
since the vibrational density of states $\rho(E)$ is 
a steeply-increasing function of the vibrational energy $E$ 
(see Fig.\,4 of Draine \& Li 2001),
and for large molecules the number of states is often 
too large to be tractable even when $E$ is small.

The only difference between the ``thermal'' approach 
and the ``statistical'' approach concerns the calculation 
of the downward transition rates $\bT_{lu}$ which, in contrast 
to the ``exact-statistical'' treatment (see \S3.2),
uses the notion of ``grain temperature'':
instead of using degeneracies $g_u$ and $g_l$,
the thermal approach replaces $g_l/g_u$ 
by $\frac{\Delta E_l}{\Delta E_u}\frac{1}{\exp(h\nu/\kB T_u)-1}$~,
where $T_u$ is the ``temperature'' of the upper level $u$;
i.e., the thermal approach assumes that the emission from
a molecule with vibrational energy $E$ and $N_a$ atoms 
can be approximated by the average emission of a system 
containing $3(N_a-2)$ vibrating harmonic oscillators,
each with a fundamental frequency $\nu_j$, 
connected to a heat bath with temperature $T$
such that the expectation value of the energy 
of the vibrational system would be 
$E(T)= \sum_{j=1}^{N_m}\frac{h\nu_j}{\exp(h\nu_j/\kB T)-1}$.
When the number of modes is large, this summation 
contains many terms. Therefore, we take the Debye model
discussed in \S3.2 for silicate grains and the C--C modes 
of PAHs to estimate the grain ``temperature'' associated
with internal energy $E$.

So far, the grains are assumed to undergo ``discrete cooling'';
i.e., the grains make discrete transitions to energy levels 
$l<u$ by emission of single photons.
There are substantial computational advantages 
(Guhathakurta \& Draine 1989; Draine \& Li 2001)
if the cooling of the grains is approximated as 
continuous rather than discrete,
so that the only downward transition from a level $u$ 
is to the adjacent level $u-1$
(i.e. $\bT_{lu} = 0$ for $u-l>1$).
We refer to this as the ``thermal continuous'' cooling approximation.

In Figure 4 we also plot the {\it cumulative} energy probability 
distributions for selected PAHs excited by the MMP ISRF
obtained from the thermal-discrete model, 
and the thermal-continuous model. 
It is seen that the energy distributions $P(E)$ 
found using the thermal-discrete model 
and the thermal-continuous model are both
in good overall agreement with the results 
of the exact-statistical calculation.
The IR emission spectra obtained from the thermal-discrete model,
as shown in Figure 5, are almost identical to those of 
the exact-statistical model.
The thermal-continuous cooling model also results in spectra 
which are very close to those computed using the exact-statistical 
model (not shown here; see Figs.\,14,15 of Draine \& Li 2001).

\begin{figure}[]  
\vspace*{-1.5em}
\centerline{
\psfig{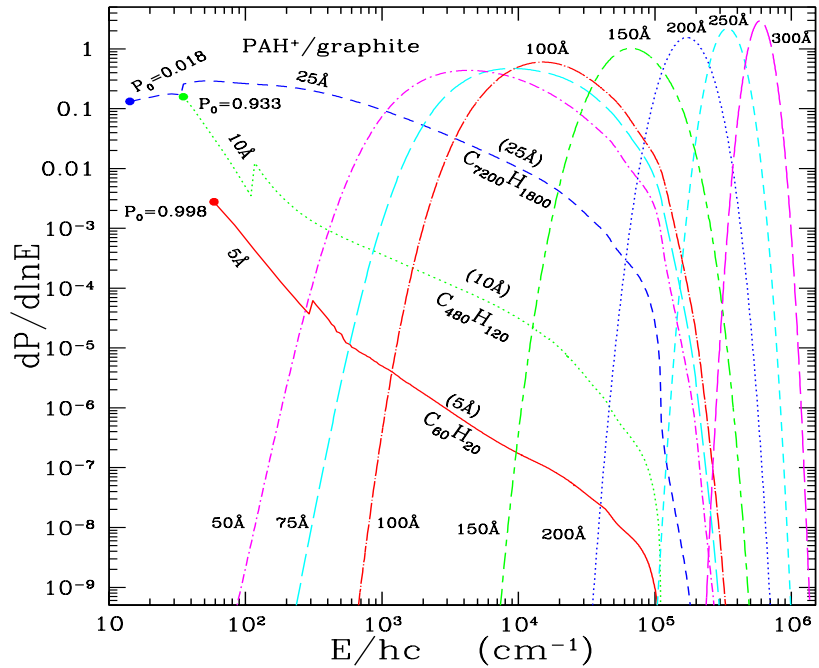}}
\vspace*{-0.8em}
\caption{
 	\footnotesize
        The energy probability distribution functions 
        for charged carbonaceous grains 
        ($a=5\Angstrom$ [C$_{60}$H$_{20}$], 
         10$\Angstrom$ [C$_{480}$H$_{120}$], 
         25$\Angstrom$ [C$_{7200}$H$_{1800}$], 
         50, 75, 100, 150, 200, 250, 300\,\AA)
	illuminated by the general ISRF.
        The discontinuity in the 5, 10, and 25$\Angstrom$ curves 
	is due to the change of the estimate 
        for grain vibrational ``temperature'' at the 20th vibrational
        mode (see Draine \& Li 2001). For 5, 10, and 25\AA\ a dot
        indicates the first excited state, and $P_0$ is
        the probability of being in the ground state.
        Taken from Li \& Draine (2001b).
        }
\vspace*{-1.5em}
\end{figure}

Figure 6 shows the energy probability distribution 
functions found for PAHs with radii
$a=5, 10, 25, 50, 75, 100, 150, 200, 300$\AA\ 
illuminated by the general ISRF.
It is seen that very small grains ($a$$\ltsim$$100\Angstrom$) 
have a very broad $P(E)$, and the smallest grains 
($a$$<$$30\Angstrom$) have an appreciable probability $P_0$
of being found in the vibrational ground state $E=0$.
As the grain size increases, $P(E)$ becomes narrower, 
so that it can be approximated by 
a delta function for $a$$>$$250\Angstrom$. 
However, for radii as large as $a$=$200\Angstrom$, 
grains have energy distribution functions 
which are broad enough that the emission spectrum 
deviates noticeably from the emission spectrum for 
grains at a single ``steady-state'' temperature $T$, 
as shown in Figure 7. For accurate computation of 
IR emission spectra it is therefore important to 
properly calculate the energy distribution function $P(E)$, 
including grain sizes which are large enough that the average 
thermal energy content exceeds a few eV.

\begin{figure}[]			
\vspace*{-1em}
\centerline{
\psfig{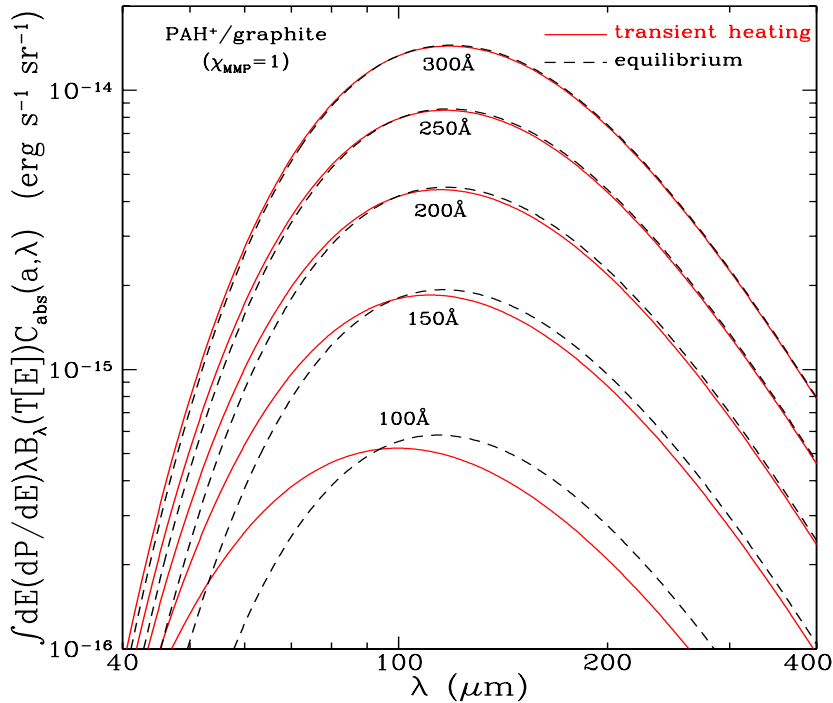}}
\vspace*{-0.5em}
\caption{
	\footnotesize
	Infrared emission spectra for small carbonaceous grains
	of various sizes heated by the general ISRF,
	calculated using the full energy distribution function 
        $P(E)$ (solid lines); also shown (broken lines) are spectra
	computed for grains at the ``equilibrium'' temperature $T$. 
	Transient heating effects lead to
        significantly more short wavelength emission for 
        $a \ltsim 200\Angstrom$. Taken from Li \& Draine (2001b).
	}
\vspace*{-1.0em}
\end{figure}

\vspace*{-1.0em}
\section{Excitation of Photoluminescence}
\vspace*{-0.3em}
Many materials are capable of emitting visible luminescence 
when subjected to some form of excitation such as
UV light (photoluminescence),
nuclear radiation such as $\gamma$ rays and $\alpha$
and $\beta$ particles (scintillation), 
mechanical shock (triboluminescence),
heat (thermoluminescence),
chemical reactions (chemiluminescence),
and electric fields (electroluminescence).
In this section we will restrict our attention
to photoluminescence, since the interstellar
dust luminescence, as manifested by the Extended
Red Emission ubiquitously seen in interstellar
environments, is believed to be a photon-driven 
process (Witt \& Schild 1985; Duley 1985; 
Smith \& Witt 2001; Witt \& Vijh 2004). 

There are two pre-requisites for luminescence:
(1) the luminescent material must have a semiconductor 
structure with a nonzero band gap $\Eg$
(e.g. metals do not luminesce since they have no band gap);
(2) energy must be imparted to this material before
luminescence can take place.
The mechanism of photoluminescence in semiconductors 
is schematically illustrated in Figure 8, which plots
the $E$--$k$ diagrams for a direct band gap material
({\it left}) and an indirect gap material ({\it right}),
where $E$ and $k$ are respectively the kinetic energy
and wave vector (or ``momentum vector'') of the electron 
or hole ($E=k^2\hbar^2/2m_\ast$, where $\hbar$$\equiv$$h/2\pi$ 
is the Planck constant $h$ divided by 2$\pi$, 
and $m_\ast$ is the electron or hole effective mass). 
The direct and indirect gap materials
are distinguished by their relative positions 
of the conduction band minimum and the valence band maximum
in the Brillouin zone (the volume of $k$ space containing 
all the values of $k$ up to $\pi/a$ where $a$ is the unit lattice 
cell dimension). 
In a direct gap material, both the conduction band 
minimum and the valence band maximum occur at the zone 
center where $k=0$. 
In an indirect gap material, however, the conduction band minimum
does not occur at $k=0$, but rather at some other values of $k$
which is usually at the zone edge or close to it (see Fox 2001).

Upon absorption of an UV or visible photon with an energy 
$\hbar\omega_{\rm exc}$ exceeding the band gap $\Eg$
(the gap in energy between the valence band and 
the conduction band) of the material, an electron-hole 
pair is created and the electron (hole) is excited to
states high up in the conduction (valence) band
(see Fig.\,8).

During a photon absorption process in semiconductors, 
we must conserve both energy and momentum.
In a direct band gap material, the conduction band 
minimum and the valence band maximum have the same $k$ values
(i.e., $\hbar\bK_i = \hbar\bK_f$, where $\bK_i$ and 
$\bK_f$ are respectively the wave vectors 
of the initial and final electron states;
this implies that the electron wave vector should not 
change significantly during a photon absorption process),
conservation of momentum is guaranteed for
the photoexcitation of the electron which only 
involves a UV or visible photon: 
$\hbar\bK_i + \hbar\bKpht \approx \hbar\bK_i = \hbar\bK_f$, 
since $\bKpht$, the wave vector of the absorbed photon
(which is in the order of $2\pi/\lambda \sim 10^{5}\cm^{-1}$),
is negligible compared to the electron wave vector
(which is related to the size of the Brillouin zone 
$\pi/a\sim 10^{8}\cm^{-1}$, where the unit cell dimension $a$ 
is in the order of a few angstroms). 
This implies that in a direct band gap material,
the electron wave vector does not change significantly 
during a photon absorption process. 
We therefore represent photon absorption processes 
by vertical arrows in the $E$--$k$ diagrams (see Fig.\,8).

In contrast, for an indirect band gap material of which 
the conduction band minimum and the valence band maximum 
have different $k$ values (see Fig.\,8), 
conservation of momentum implies that 
the photon absorption process must be assisted 
by either absorbing or emitting a phonon
(a quantum of lattice vibration),
because the electron wave vector must change significantly 
in jumping from the valence band in state ($E_i$,$\bK_i$) 
to a state ($E_f$,$\bK_f$) in the conduction band,
and the absorption of a photon alone
can not provide the required momentum change
since $|\bKpht|\ll |\bK_i - \bK_f|$.


The excited electron and hole will not remain
in their initial excited states for very long; instead,
they will relax very rapidly ($\simali$$10^{-13}\s$)
to the lowest energy states within their respective bands 
by emitting phonons. When the electron (hole) finally
arrives at the bottom (top) of the conduction (valence)
band, the electron-hole pair can recombine {\it radiatively}
with the emission of a photon (luminescence),
or {\it nonradiatively} by transferring the electron's 
energy to impurities or defects in the material 
or dangling bonds at the surface.

Just like the photon {\it absorption} process discussed above,
the electron-hole recombination in a direct band gap material 
does not involve any phonons since there is no need for momentum 
change for the electron. In contrast, in an indirect gap material, 
the excited electron located in the conduction band needs to 
undergo a change in momentum state before it can recombine with 
a hole in the valence band; conservation of momentum demands 
that the electron-hole recombination must be accompanied 
by the emission of a phonon, 
since it is not possible to make this recombination 
by the emission of a photon alone.
Compared to the photon {\it absorption} process 
in an indirect gap material for which conservation of
momentum can be fulfilled by either absorption or emission of 
a phonon, in the electron-hole {\it recombination} process
phonon absorption becomes negligible, 
whereas phonon emission becomes the dominant momentum 
conservation mediator because 
(1) the number of phonons available for absorption is
small and is rapidly decreasing at lower temperatures,
whereas the emission of phonons by electrons which are
already at a high-energy state is very probable; 
and (2) an optical transition assisted by phonon
emission occurs at a lower photon energy $\Eg-h\nuphn$
than the gap energy, whereas phonon absorption results 
in a higher photon energy of at least $\Eg+h\nuphn$, 
which can be more readily re-absorbed 
by the semiconductor nanoparticle.
But we note that the energy of a phonon ($h\nuphn$) is 
just in the order of $\simali$$0.01\eV$, much smaller than 
the energy of the electron-hole recombination luminescence photon.
Also because prior to the recombination, 
the electrons and holes respectively accumulate 
at the bottom of the conduction band 
and the top of the valence band, 
the energy separation between the electrons and the holes 
approximately equals to the energy 
of the band gap. Therefore, the luminescence emitted by 
both types of semiconductors occurs at an energy close to 
the band gap $\Eg$. 

The PL efficiency is determined by the competition
between radiative and nonradiative recombination.
For an indirect gap material, the PL process, which
requires a change in both energy and momentum for
the excited electron and hence involves both a photon 
and a phonon, is a second-order process 
with a long radiative lifetime ($\simali$$10^{-5}-10^{-3}\s$),
and therefore a relatively small efficiency
because of the competition with nonradiative combination. 
In contrast, in a direct gap material, the emission of
a PL photon does not need the assistance of a phonon 
to conserve momentum. Therefore, the PL process in
a direct gap material is a first-order process with
a much shorter radiative lifetime ($\simali$$10^{-9}-10^{-8}\s$)
and a much higher PL efficiency
in comparison with an indirect gap material. 

\begin{figure}[]			
\vspace*{-2.0em}
\hspace*{-0.5em}
\centerline{
\psfig{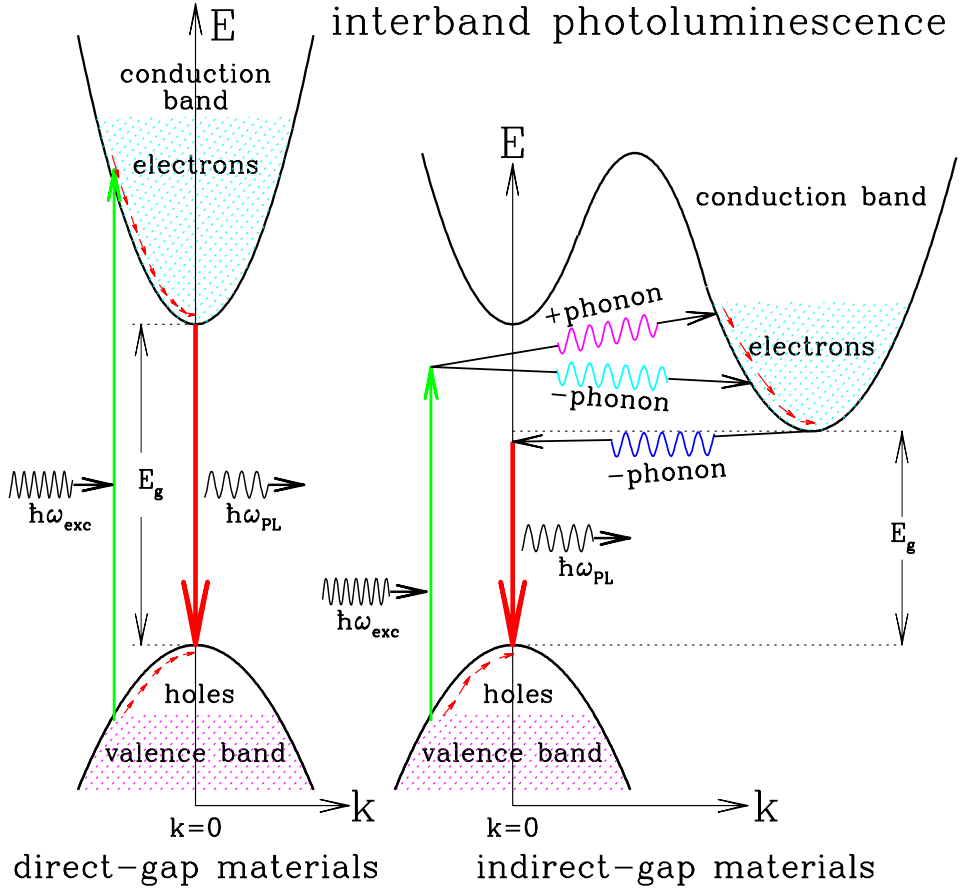}}
\vspace*{-3.0em}
\caption{
        \footnotesize
        Schematic band diagrams for the photoluminescence processes 
        in a direct gap material ({\it left}) 
        and an indirect gap material ({\it right}).
        The shaded states at the bottom of the conduction band 
        and the empty states at the top of the valence band
        respectively represent the electrons and holes
        created by the absorption of photons with energy
        $\hbar\omega_{\rm exc}$$>$$\Eg$.
        The cascade of transitions within 
        the conduction and valence bands 
        represents the rapid thermalization 
        of the excited electrons and holes
        through phonon emission.
        In a direct gap material ({\it left}), 
        the conduction band minimum and the valence band maximum 
        occur at the same $k$ values.
        Both the photon absorption and emission 
        (i.e. the electron-hole recombination)
        processes can conserve momentum without the assistance 
        of phonons, since the momentum of the absorbed or emitted
        photon is negligible compared to the momentum of the electron.
        We therefore represent photon absorption and emission 
        processes by vertical arrows on $E$--$k$ diagrams.
        In an indirect gap material ({\it right}), 
        the conduction band minimum
        and the valence band maximum occur at different $k$ values.
        As a result, to conserve momentum, the photon absorption 
        process must involve either absorption 
        (indicated by a ``$+$'' sign)
        or emission (indicated by a ``$-$'' sign) of a phonon, 
        while the PL process requires the emission of a phonon.
        Since the energy of a phonon ($\simali$$0.01\eV$)
        is much smaller than the energy of the PL photon,
        for an indirect gap material, the peak energy of 
        the PL also roughly reflects its band gap.
        }
\end{figure}

\vspace*{0.4em}
However, for particles in the nanometer size domain, 
we would expect substantial changes in 
both the efficiency and the peak energy of 
the photoluminescence due to the quantum confinement effect. 
This can be understood in terms of the Heisenberg uncertainty 
principle. Unlike in bulk materials the electrons and holes
are free to move within their respective bands in all three
directions, in nanoparticles the electrons and holes are 
spatially confined and hence their motion is quantized in
all three dimensions.
The spatial confinement of a particle of mass $m$ 
to a region in a given direction (say, along the $x$ axis) 
of length $\Delta x$ would introduce an uncertainty in 
its momentum $\Delta p_x \sim \hbar/\Delta x$ 
and increase its kinetic energy by an amount
$E_{\rm QC} \sim \left(\Delta p_x\right)^2/2m 
\sim \hbar^2/2m\left(\Delta x\right)^2$.
A simple particle-in-a-box analysis,
using the Schr\"{o}dinger's equation and the effective
mass approximation, shows that 
the ground state quantum confinement energy would be
$\EQC \sim \frac{3\hbar^2}{8\,\meff} \left(\frac{\pi}{a}\right)^2$, 
where $\meff \equiv \meeff\mheff/\left(\meeff+\mheff\right)$ 
is the reduced effective mass of the electron-hole pair
($\meeff$ and $\mheff$ are respectively the effective mass 
of the electron and hole) (Fox 2001). 
For nano-sized particles, the quantum confinement
effect becomes significant since the confinement energy 
$\EQC$ would be comparable to or greater than 
their thermal energy $\Eth \sim \frac{3}{2}\kB T$
at the temperature range expected for nanoparticles 
in the diffuse ISM (see Figs.\,7,8 of Draine \& Li 2001 
and Fig.\,3 of Li \& Draine 2002a).
For example, a silicon grain with $\meeff \approx 0.98\me$
and $\mheff \approx 0.52\me$ ($\me$ is the free electron mass)
smaller than $\simali$8$\nm$
would exhibit quantum effects at a temperature 
$T$$\simali$100$\K$ (which is expected for 
nano-sized silicon dust in the diffuse ISM; 
see Fig.\,3 of Li \& Draine 2002a)
with $\EQC \approx 0.83\left(a/{\rm nm}\right)^{-2}\eV
> \Eth\approx 0.013\left(T/100\K\right)\eV$.  
More detailed studies show that silicon nanocrystals
exhibit an $a^{-1.39}$ gap-size dependence:
$\Eg\approx E_0 + 1.42\left(a/{\rm nm}\right)^{-1.39}\eV$,
where $E_0\approx 1.17\eV$ is the bulk silicon band gap
(Delerue, Allan, \& Lannoo 1993).
For nanodiamonds, an $a^{-2}$ gap-size dependence
was derived from the X-ray absorption 
spectrum measurements (Chang et al.\ 1999; 
also see Raty et al.\ 2003): 
$\Eg\approx E_0 + 0.38\left(a/{\rm nm}\right)^{-2}\eV$,
where $E_0\approx 5.47\eV$ is the bulk diamond band gap.

It is apparent, therefore, the quantum confinement
effect would lead to a progressive widening of the
band gap of a nano-sized semiconductor as its size 
is reduced, along with a broadening of the electron-hole 
pair state in momentum space (i.e. an increased overlap
between the electron and hole wavefunctions),
and a decreasing probability for the pair to find
a nonradiative recombination center, 
provided that the surface dangling bonds are passivated
which would otherwise act as traps for the carriers
and quench the PL.
While the former would shift the PL peak 
to higher energies, the latter two effects would 
greatly enhance the electron-hole radiative recombination 
probability and result in a higher PL efficiency.
This is one of the main reasons why silicon nanocrystals 
are proposed recently by Witt and his coworkers 
(Witt et al.\ 1998, Smith \& Witt 2001) 
and by Ledoux et al.\ (1998, 2001) as the ERE carrier 
since they are capable of luminescing very efficiently 
(with essentially 100\% quantum efficiency)
in the energy range ($\simali$1.4--2.4$\eV$)
over which the ERE has been observed in astronomical sources,
while bulk silicon, an indirect gap semiconductor 
with a band gap of $\simali$1.17$\eV$ at $T=0\K$, 
does not luminesce.

\vspace*{-1.2em}
\section{An Overview of Interstellar Nanoparticle Species}
\vspace*{-0.5em}
\subsection{Nano Carbon Grains: Polycyclic Aromatic Hydrocarbon Dust}
\vspace*{-0.3em}
As the most abundant interstellar nanoparticle species,
nano carbon grains (mainly PAHs), containing $\simali$15\%
of the interstellar C abundance (Li \& Draine 2001b),
reveal their presence in the ISM by emitting a prominent 
set of ``UIR'' bands at 3.3, 6.2, 7.7, 8.6, 
and 11.3$\mum$ (see \S1 and \S2).
Modern research on astrophysical PAHs started
in the mid 1980s with the seminal studies of L\'eger \& Puget
(1984) and Allamandola et al.\ (1985) who by the first time
explicitly proposed PAH molecules as the ``UIR'' band carrier.
The PAH model is now gaining increasing acceptance because of 
(1) the close resemblance of the ``UIR'' spectra (frequencies and 
relative intensities) to the vibrational spectra of PAH molecules 
(see Allamandola \& Hudgins 2003 for a recent review);
(2) the ability of a PAH molecule to emit efficiently 
in the ``UIR'' wavelength range following single photon heating 
(L\'{e}ger \& Puget 1984; Allamandola et al.\ 1985, 1989; 
Draine \& Li 2001; also see \S3);
and (3) the success of the PAH model in quantitatively
reproducing the observed mid-IR spectra 
of the Milky Way diffuse ISM (Li \& Draine 2001b), 
the quiescent molecular cloud SMC B1\#1 
in the Small Magellanic Cloud (Li \& Draine 2002c),
and the ``UIR'' band ratios for a wide range of environments 
ranging from reflection nebulae, HII regions, 
photodissociation regions,
molecular clouds in the Milky Galaxy to normal galaxies,
starburst galaxies, and a Seyfert 2 galaxy (Draine \& Li 2001). 

Recently, the PAH model further gained its strength 
from a close fit to the observed ``UIR'' bands of $\vdb$, 
a UV-poor reflection nebula, which was considered as 
one of the major challenges to identification of 
the ``UIR'' bands with PAH molecules
(see Uchida, Sellgren, \& Werner 1998),
since small, neutral PAH molecules 
have little or no absorption at visible wavelengths 
and therefore require UV photons for excitation.
Li \& Draine (2002b) have shown that 
the ``astronomical'' PAH model, 
incorporating the experimental result that 
the visual absorption edge shifts to 
longer wavelength upon ionization and/or 
as the PAH size increases 
(see Appendix A2 of Li \& Draine 2002b and references therein),
can closely reproduce the observed IR emission bands of
this reflection nebula, 
and is also able to account for the observed dependence 
of the 12$\mum$ IRAS emission on the effective temperature 
of the illuminating star (Sellgren, Luan, \& Werner 1990).

Henning \& Schnaiter (1998) suggested that nano-sized
hydrogenated carbon grains may be responsible for 
the 2175$\Angstrom$ extinction hump as well as 
the 3.4$\mum$ C--H absorption feature. 
But this seems to be in conflict with
the detection of the 3.4$\mum$ feature in regions
where the 2175$\Angstrom$ hump is not seen, 
e.g., along the sightline toward HD\,204827 (Valencic et al.\ 2003)
and the Taurus cloud (Whittet et al.\ 2003). 

Kroto et al.\ (1985) first proposed that C$_{60}$
could be present in the ISM with a considerable
quantity. This molecule and its related species 
have later been proposed as the carriers of 
the 2175$\Angstrom$ extinction hump, the diffuse
interstellar bands (DIBs), the ``UIR'' bands, 
and the ERE (see Webster 1993 and references therein).
Foing \& Ehrenfreund (1994) attributed the two
DIBs at 9577$\Angstrom$ and 9632$\Angstrom$ to C$_{60}^{+}$.
However, attempts to search for these molecules in
the UV and IR were unsuccessful (Snow \& Seab 1989; 
Somerville \& Bellis 1989; Moutou et al.\ 1999; Herbig 2000).
These molecules were estimated to consume at most 
$<$0.7$\ppm$ carbon (Moutou et al.\ 1999). 
Therefore, C$_{60}$ is at most a minor 
component of the interstellar dust family. 

\vspace*{-0.5em}
\subsection{Nano Silicate Grains}
\vspace*{-0.2em}
If nano silicate grains are present in the ISM,
single-photon heating by starlight (see \S3)
will cause these grains to radiate in the 10$\micron$ feature
(see, e.g., Draine \& Anderson 1985, Li \& Draine 2001a).
The absence of a 10$\micron$ emission feature in IRAS spectra 
led D\'esert et al.\ (1986) to conclude that not more than 
1\% of Si could be in $a$$<$$15\Angstrom$ silicate grains, 
and on this basis recent grain models
(see Li \& Draine 2001a and references therein)
have excluded silicate nanoparticles as a significant grain component.
Nondetection of the 10$\mum$ silicate emission feature 
in the diffuse ISM by ISO (Mattila et al.\ 1996)
and the IRTS (Onaka et al.\ 1996) 
appeared to confirm this conclusion.

However, the presence of nano silicate grains can not be ruled out 
since the 10$\mum$ silicate emission feature may be hidden by 
the dominant PAH features.
Li \& Draine (2001a) have placed quantitative upper limits on
the abundances of both amorphous and crystalline 
silicate nanoparticles by calculating spectra 
for such tiny grains heated by starlight, 
and comparing to measurements of the IR emission
of the diffuse ISM by 
IRTS (Onaka et al.\ 1996) and by
the DIRBE instrument on the COBE satellite.
The interstellar extinction curve and the 10$\mum$ silicate
absorption profile are also invoked to provide further constraints.
It was found that, contrary to previous work, 
as much as $\sim$10\% of interstellar Si could be 
in $a$$\ltsim$$15\Angstrom$ silicate grains
without violating observational constraints,
and not more than $\sim$5\% of the Si can be 
in crystalline silicates (of any size).

\vspace*{-0.5em}
\subsection{Nano Silicon Grains}
\vspace*{-0.2em}
The presence of a population of silicon nanoparticles
containing $\ltsim$$5\%$ of the total interstellar 
dust mass (with Si/H $\approx6\ppm$) in the ISM 
was proposed by Witt and his coworkers (Witt et al.\ 1998, 
Smith \& Witt 2002) and Ledoux et al.\ (1998) to account
for the ERE phenomenon.
Witt et al.\ (1998) suggested that the formation of 
interstellar SNPs could occur as a result of the nucleation 
of SiO molecules in oxygen-rich stellar outflows,
followed by annealing and phase separation into an elemental
silicon phase in the core and a passivating mantle of SiO$_2$. 
The SiO$_2$ mantle is crucial for the SNP model 
since SNPs luminesce efficiently only when their
surface dangling Si bonds are passivated.

The SNP model is recently receiving much attention because
(1) as a consequence of quantum confinement (see \S4),
the SNP photoluminescence provides so far the best match 
to the observed ERE in both spectral profile (i.e. peak position 
and width) and the required extremely high quantum efficiency;
(2) as shown by Smith \& Witt (2002) in terms of 
a SNP photoionization and photofragmentation theory
together with the experimentally established fact that
photoionization quenches the PL of SNPs,
this model successfully reproduces the observed 
dependences of the ERE intensity, the ERE quantum efficiency, 
and the ERE peak wavelength on the intensity and 
hardness of the UV/visible radiation fields of 
a wide variety of dusty environments;
(3) as already mentioned in \S2, other candidate materials
all fail to simultaneously satisfy 
the ERE spectral characteristics and the 
high quantum efficiency requirement;
it is worth noting that HAC, the long-thought ERE candidate,
also falls in this category: 
HAC luminesces efficiently only in the near UV
(quantum efficiencies as high as 10\% have been
reported; see Rusli, Robertson, \& Amaratunga 1996 
and references therein) when it has a higher degree 
of hydrogenation (indicating a larger band gap), 
dehydrogenation is expected to reduce the band gap 
and thus can redshift the PL peak into the wavelength range 
where the ERE is observed, but this results in
an exponential drop in its PL efficiency by
a factor of $\simali$$10^4$.

As already mentioned in \S2, although the SNP model 
seems very attractive, it also has some weakness:
it has been shown by Li \& Draine (2002a) that
SNPs with oxide coatings, if they are free-flying in the ISM,
would emit strongly in the 20$\mum$ O-Si-O bending band. 
Existing COBE-DIRBE 25$\micron$ photometry appears to 
already rule out such high abundances of SNPs.
If SNPs are responsible for the ERE from the diffuse ISM,
they must either be in $a$$\gtsim$$50\Angstrom$ clusters, 
or attached to larger ``host'' grains.
This problem can not be resolved by invoking hydrogen passivation 
of SNPs because H-passivated SNPs luminesce at blue and 
near-UV wavelengths (Zhou, Brus, \& Friesner 2003)
while blue PL is not observed under interstellar conditions
(Rush \& Witt 1975; Vijh, Witt, \& Gordon 2003). Very recently, 
Witt \& Vijh (2004) argued that Fe- or C-passivated SNPs 
could overcome this problem. It would be very helpful to 
perform detailed calculations of the 11.3$\mum$ Si--C emission
feature of C-passivated SNPs and compare with the upcoming
SIRTF ({\it Space Infrared Telescope Facility}) high resolution
spectra of the diffuse ISM. 
Previously, Whittet, Duley, \& Martin (1990) deduced that 
the abundance of Si in SiC dust is no more than $\simali$5\% 
of that in silicates, through an analysis of the 7.5--13.5$\mum$ 
absorption spectra of 10 Galactic Center sources.
It would also be very useful to investigate the IR emission
properties of Fe-passivated SNPs which luminesce even more
efficiently than O-passivated SNPs in the wavelength range 
over which the ERE is observed (Mavi et al.\ 2003).

\vspace*{-0.3em}
\subsection{Nano Diamond Grains}
\vspace*{-0.1em}
Cosmic nanodiamonds were first detected in primitive carbonaceous 
meteorites and identified as presolar in origin based on 
their isotopic anomalies (Lewis et al.\ 1987), although 
their presence in the ISM was proposed almost two decades 
earlier by Saslaw \& Gaustad (1969)
to explain the interstellar UV extinction curve.
Five years later, Allamandola et al.\ (1992)
attributed the 3.47$\mum$ absorption band seen toward
a large number of protostars to the tertiary C--H stretching
mode in diamond-like carbonaceous materials.
Jones \& d'Hendecourt (2000) further suggested that
surface-reconstructed (to $sp^2$-bonded carbon) nanodiamonds
could be responsible for the ``UIR'' emission features,
the 2175$\Angstrom$ extinction hump (also see Sandford 1996),
and a part of the far-UV extinction
at $\lambda^{-1}\gtsim 7\mum^{-1}$.
More recently, circumstellar nanodiamonds were identified in
the dust disks or envelopes surrounding
two Herbig Ae/Be stars HD 97048 and Elias 1
and one post-asymptotic giant branch (AGB) star HR 4049,
based on the 3.43$\mum$ and 3.53$\mum$ 
C--H stretching emission features 
expected for surface-hydrogenated nanodiamonds
(Guillois, Ledoux, \& Reynaud 1999; 
van Kerckhoven, Tielens, \& Waelkens 2002). 

Presolar meteoritic nanodiamonds 
were found to have a log-normal size distribution
with a median radius $\simali$1.3$\nm$
(Lewis, Anders, \& Draine 1989)
and an abundance as much as $\simali$0.1\% 
of the total mass in some primitive meteorites,
more abundant than any other presolar grains
by over two orders of magnitude. 
However, it is important to recognize that this abundance 
may not be representative of its original ISM proportion
but rather its durability (see Draine 2003a).
Indeed, Lewis et al.\ (1989) found that
in interstellar space, as much as 10\% of 
the interstellar carbon ($\simali$36$\ppm$) 
could be in the form of nanodiamonds 
without violating the constraints placed by
the interstellar extinction curve. 
However, a much more stringent upper limit of $\simali$0.1$\ppm$
was derived by Tielens et al.\ (2000) based on 
nondetection of the characteristic 3.43$\mum$ 
and 3.53$\mum$ C--H stretching emission features
in the ISM. Of course, interstellar nanodiamonds could 
be more abundant if the bulk of them are not hydrogenated. 

\vspace*{-0.3em}
\subsection{Nano Titanium Carbide Grains}
\vspace*{-0.1em}
Presolar titanium carbide (TiC) grains were first
identified in primitive meteorites as nano-sized
inclusions embedded in micrometer-sized presolar graphite 
grains (Bernatowicz et al.\ 1996). 
Very recently, von Helden et al.\ (2000) proposed that
TiC nanocrystals could be responsible for the prominent
21$\mum$ emission feature detected in over a dozen 
carbon-rich post-AGB stars. 
This mysterious 21$\mum$ feature still remains unidentified 
since its first detection in the IRAS LRS ({\it Low Resolution 
Spectrometer}) spectra of four post-AGB stars (Kwok, Volk,
\& Hrivnak 1989). While bulk TiC samples show no resonance 
around 21$\mum$ (Henning \& Mutschke 2001),
laboratory absorption spectra of TiC nanocrystal clusters 
exhibit a strong 21$\mum$ feature, closely resembling 
the observed 21$\mum$ feature both in position, width and 
in spectral detail (von Helden et al.\ 2000). 

We have examined the nano-TiC proposal by comparing 
the maximum {\it available} with the minimum {\it required} 
TiC dust mass inferred from the nano-TiC model (see Li 2003). 
It is found that if the UV/visible absorption properties 
of TiC nanograins are like their bulk counterparts, 
the model-required TiC dust mass would exceed 
the maximum available TiC mass
by over two orders of magnitude
(also see Chigai et al.\ 2003; Hony et al.\ 2003).  
One may argue that nano TiC might have a much higher
UV/visible absorptivity so that the available TiC mass 
may be sufficient to account for the observed 21$\mum$ 
emission feature.
However, the Kramers-Kronig dispersion relations, 
which relate the wavelength-integrated extinction cross sections 
to the total dust mass (Purcell 1969; Draine 2003b),
would impose a lower bound on the TiC mass. 
We have shown that this Kramers-Kronig lower limit exceeds 
the maximum available TiC mass by a factor of at least 
$\simali$50 (see Li 2003 for details). 
Therefore, it is unlikely that TiC nanoparticles 
are responsible for the 21$\mum$ emission feature 
of post-AGB stars.

\vspace*{-1.0em}
\section{Concluding Remarks}
\vspace*{-0.5em}
Interstellar nanoparticles are important 
as emitters of near and mid IR radiation,
as absorbers of far-UV radiation, 
as heating agents of the interstellar gas,
and possibly as luminescing agents of red light.
Our understanding of interstellar nanoparticles
obviously remains incomplete.
Besides the long-standing problems regarding the carriers 
of the 2175$\Angstrom$ extinction hump and the ERE,
there are many unanswered questions which will be demanding 
close attention in the future:
\begin{itemize}
\vspace*{-0.5em}
\item The origin and evolution of interstellar PAHs are 
      not very clear. Suggested sources for interstellar
      PAHs include (1) injection (into the ISM) of PAHs 
      formed in carbon star outflows (Latter 1991); 
      (2) shattering of carbonaceous interstellar dust 
      or of photoprocessed interstellar dust 
      organic mantles (Greenberg et al.\ 2000) 
      by grain-grain collisions 
      in interstellar shocks (Jones, Tielens, \& Hollenbach 1996);
      (3) {\it in-situ} formation through ion-molecule
      reactions (Herbst 1991). 
      Although interstellar PAHs containing more than
      $\simali$20--30 carbon atoms can survive the UV
      radiation field (Guhathakurta \& Draine 1989; 
      Jochims et al.\ 1994; Allain, Leach, \& Sedlmayr 1996;
      Le Page, Snow, \& Bierbaum 2003),
      they are still subject to destruction by sputtering
      in interstellar shock waves 
      (but also see Tielens et al.\ [2000] who argued that
       destruction of PAHs by sputtering is unimportant, 
       except in extreme environments such as very young
       supernova remnants with $\gtsim$200$\kms$ shocks,
       because they couple very well dynamically to 
       the gas which cools down rapidly),
      chemical attack by atomic oxygen,
      and coagulation in dense regions (Draine 1994).
      Clearly, a detailed study of the evolution of
      interstellar PAHs would be very valuable.
\item The physics and astrophysics of the radiative electronic 
      transitions of interstellar PAHs are not fully understood.

      Small neutral PAHs with $\ltsim$40 C atoms 
      are expected to emit near-UV and blue photons 
      through fluorescence, phosphorescence,
      and perhaps also through the recurrent Poincar\'e fluorescence.
      Laboratory studies show that the fluorescence
      quantum yield can be quite high for isolated
      molecules (e.g., fluorescence quantum yields 
      in the range of 10\%--45\% have been measured   
      for gas-phase, collision-free naphthalene
      C$_{10}$H$_{8}$ [Reyl\'e \& Br\'echignac 2000]).
      However, no PL shortward of $\lambda$=$5000\Angstrom$ 
      has been seen in the ISM
      (Rush \& Witt 1975; Vijh, Witt, \& Gordon 2003).
      Although PAH ions do not exhibit strong ordinary
      fluorescence or phosphorescence (see the caption of Fig.\,2), 
      they can undergo Poincar\'e fluorescence 
      which may have a quantum yield larger than one. 
      Therefore, it is difficult to explain
      nondetection of blue PL in the ISM in terms of 
      PAH photoionization. 
      Moreover, for PAHs containing $\ltsim$40 C atoms
      in the diffuse ISM, the probability of finding them 
      in a nonzero charge state is smaller than $\simali$30\% 
      (see Fig.\,7 of Li \& Draine 2001b).
      One may argue that interstellar PAHs are larger
      (which is true, as shown in Li \& Draine 2001b,
       the average [diffuse ISM] PAH size is $\simali$6$\Angstrom$,
       corresponding to $\simali$100 C atoms)
      so that their fluorescence mainly occurs at longer
      wavelengths, say, in the wavelength range over which
      the ERE is observed. 

      However, if interstellar PAHs
      indeed luminesce in the ERE band with the required 
      quantum efficiency ($\simali$100\%; see \S2), 
      the ``UIR'' emission bands expected for these 
      luminescing PAHs would be strongly suppressed 
      because a considerable fraction of the excitation 
      energy is released in the form of PL photons
      (e.g., in the diffuse ISM with a mean energy 
       $\approx$\,2.1$\eV$ for ERE photons 
       [Szomoru \& Guhathakurta 1998],
       only $\simali$60\% of the original excitation
       energy is available as heat for PAHs which have
       a mean absorbed photon energy $\approx$5.2$\eV$
       [Draine \& Li 2001];
       this could be even worse since the Poincar\'e fluorescence 
       may have a quantum yield higher than one).
       One consequence of this would be that even more 
       carbon atoms in the form of PAHs would be needed to 
       explain the intensity of the ``UIR'' bands,
       deteriorating the already tightened carbon budget
       ``crisis'' (Snow \& Witt 1996).
\vspace*{-0.4em}
\item It is unclear how different the optical and thermal 
      properties of nanometer-sized materials are
      compared with their bulk counterparts,
      although it is generally believed that they
      may be very different. 
\begin{itemize}
\vspace*{-0.4em}
\item For a small metallic grain, the imaginary part
      $\ei$ of its dielectric function 
      $\epsilon(\lambda) = \er +i\,\ei$
      is expected to be larger compared to that of its
      bulk counterpart, as a consequence of the so-called 
      {\it electron mean free path limitation} effect
      (e.g. see Bohren \& Huffman 1983).
      This is easier to understand if we decompose $\epsilon$
      into two components $\eb$ and $\ef$, contributed by
      bound charges (``interband transitions'')
      and free electrons, respectively.
      The free electron component is well described by
      the Drude theory 
      $\epsilon = 1 - \frac{\omega_p^2}{\omega^2+i\,\gamma\omega}$
      with an imaginary part $\ei = \frac{\gamma\omega_p^2}
      {\omega\left(\omega^2+\gamma^2\right)}$, where the plasma
      frequency $\omega_p$ is related to 
      the free electron density $n_e$:
      $\omega_p^2 = n_e e^2/\meeff$, 
      the damping constant $\gamma$ is related to 
      the average time $\tau$ between collisions:
      $\gamma$=$1/\tau$. In bulk materials, $\gamma$ is
      mainly determined by the scattering of the electrons with
      phonons (lattice vibrations), and to a lesser degree, 
      with electrons, lattice defects, or impurities. 
      However, for particles in the nanometer size domain, 
      $\gamma$ is increased because of the additional collisions 
      of the conducting electrons with the grain boundary:
      $\gamma = \gamma_{\rm bulk} + \frac{v_F}{\beta a}$, 
      where $\gamma_{\rm bulk}$ is the bulk metal damping constant, 
      $v_F$ is the electron velocity at the Fermi surface, 
      $\beta$ is a dimensionless constant of order unity which 
      depends on the character of the scattering at the boundary: 
      $\beta$=1 for classic isotropic scattering, 
      $\beta$=4/3 for classic diffusive scattering, 
      $\beta$=1.16 or $\beta$=1.33 for scattering based on the quantum 
      particle-in-a-box model (see Coronado \& Schatz 2003 and
      references therein). Since $\omega^2\gg \gamma^2$ in metals
      near the plasma frequency, $\ei$ can be written as 
      $\ei = \ei_{\rm bulk} + \frac{v_F \omega_p^2}{\beta a \omega^3}$.
      Clearly, for a metallic grain $\ei$ increases as the grain 
      becomes smaller.\\[1mm]
      In contrast, our knowledge regarding the size dependence of
      the dielectric function of {\it dielectric} materials is 
      controversial. For example, some theoretical and experimental 
      studies have concluded that the dielectric function of 
      Si nanoparticles is significantly reduced relative to 
      the bulk value (Koshida et al.\ 1993; Wang \& Zunger 1994; 
      Tsu, Babic, \& Ioriatti 1997; Amans et al.\ 2003),
      which is generally attributed to the quantum confinement effect.
      But we find that the absorption and reflectivity measurements 
      for porous silicon appear to be consistent with bulk Si together 
      with voids and SiO$_2$ (see Li \& Draine 2002a for details).
\vspace*{-0.5em}
\item It has been reported that the specific heat of small metal 
      particles is strongly enhanced over the bulk value
      (see  Halperin 1986, Meyer et al.\ 2003 and references therein).
      For example, a progressive decreasing of Debye temperature 
      $\Theta$ with the decrease of grain size has been observed 
      for palladium (Pd):
      $\Theta\approx 273, 226, 193, 175\K$ for bulk Pd 
      and Pd particles of radius $a=42,33,15\Angstrom$, 
      respectively (Chen et al.\ 1995 and references therein).
      This has been attributed to quantum effects on the
      vibrational spectrum in small particles:
      as a grain becomes smaller, a larger fraction
      of atoms occupy surface sites which are weakly
      bound to the grain; therefore, small grains
      are expected to have a larger low-frequency mode density 
      due to the weaker bonds of the surface atoms.

      However, although an enhancement of the vibrational
      specific heat is also expected for dielectric grains,
      the degree of enhancement is unclear 
      due to the differences between the binding and 
      structural properties of dielectrics with metals. 
      For example, based on a lattice dynamical calculation, 
      Hu et al.\ (2001) have shown that the difference in 
      vibrational specific heat between Si nanocrystals and 
      the bulk is just about a few per cent.
      The specific heat of nanocrystalline diamond measured
      by Moelle et al.\ (1998) shows a close agreement
      with that of {\it bulk} diamond.  
      It is worth noting that the heat capacity of PAHs 
      can be well described by the two 2-dimensional Debye models
      of {\it bulk} graphite (together with contributions from 
      the C--H vibrational modes) (see Fig.\,2 of Draine \& Li 2001).
      Clearly, laboratory studies of the optical and 
      thermal properties of interstellar nanoparticle analogues 
      would be very valuable.
\end{itemize}
\end{itemize}

\acknowledgements
I am extremely grateful to B.T. Draine and A.N. Witt
for their invaluable advices, comments, and suggestions.
I thank the anonymous referee for his/her very helpful 
comments and suggestions. 
I also thank A.N. Witt for the great efforts he has put into 
making the ``Astrophysics of Dust Symposium'' a real success.

\begin{small}

\end{small}

\end{document}